\documentclass[pageno]{jpaper}

\usepackage{tikz}
\usepackage{amsmath}
\usepackage{pifont}
\usepackage{hyperref}
\usepackage{enumitem}
\usepackage{xcolor}

\usepackage[compact]{titlesec}

\setlist[itemize]{leftmargin=*, noitemsep, nolistsep}
\setlist[enumerate]{leftmargin=*, noitemsep}
\usepackage[normalem]{ulem}

\hypersetup{
    colorlinks=true,
    citecolor=blue,
    urlcolor=black,
    linkcolor=blue,
}

\newtoggle{showmarks}
\toggletrue{showmarks}

\iftoggle{showmarks}{

}{

}

\newcommand{\Scheduler}{Hermes~}
\newcommand{\SchedulerT}{Hermes}

\usepackage[shortcuts]{extdash} % Allow or prevent line breaks at the hyphen.
\begin{document}

\title{Practical Scheduling for Real-World Serverless Computing}

\author{Kostis Kaffes*, Neeraja J. Yadwadkar$^{\dagger}$, Christos Kozyrakis*\\
        \textit{*Stanford University}, \textit{$^{\dagger}$UT Austin/VMware Research}}
%\email{kkaffes@stanford.edu}
%\affiliation{
%    \institution{Stanford University}
%}

%\author{Neeraja J. Yadwadkar}
%\email{neeraja@cs.stanford.edu}
%\affiliation{
%    \institution{Stanford University}
%}

%\author{Christos Kozyrakis}
%\email{kozyraki@stanford.edu}
%\affiliation{
%    \institution{Stanford University}
%}

\date{}
\maketitle

\thispagestyle{empty}

%!TEX root =main.tex
%-------------------------------------------------------------------------------
\begin{abstract}
%-------------------------------------------------------------------------------
Serverless computing has seen rapid growth due to the ease-of-use and cost-efficiency it provides.
However, function scheduling, a critical component of serverless systems, has been overlooked.
In this paper, we take a first-principles approach toward designing a scheduler
that caters to the unique characteristics of serverless functions as seen
in real-world deployments. We first create a taxonomy of scheduling policies
along three dimensions. Next, we use simulation to explore the scheduling policy space for the
function characteristics in a 14-day trace of Azure functions and conclude
that frequently used features such as late binding and random load balancing
are sub-optimal for common execution time distributions and load ranges.
We use these insights to design \SchedulerT, a scheduler for serverless functions with three key characteristics. 
First, to avoid head-of-line blocking due to high function execution time variability, \Scheduler uses a combination of early binding and processor sharing for scheduling at individual worker machines. 
Second, \Scheduler uses a hybrid load balancing approach that improves consolidation at low load while employing least-loaded balancing at high load to retain high performance. 
Third, \Scheduler is both load and locality-aware, reducing the number of cold starts compared to pure load-based policies.
We implement \Scheduler for Apache OpenWhisk and demonstrate that, for the case
of the function patterns observed both in the Azure and in other real-world
traces, it achieves up to 85\% lower function slowdown and 60\% higher
throughput compared to existing policies.
\end{abstract}

% Serverless computing has seen rapid growth due to the ease-of-use and cost-effectiveness it provides.
% However, scheduling, a critical component of cloud systems, has been overlooked in the design of serverless platforms.
% In this paper, we take a first-principles approach toward designing a scheduler that caters to the unique characteristics of serverless functions.
% We study a large-scale real-world trace and create a taxonomy of possible scheduling policies.
% We explore the policy space, and draw conclusions that guide the design of \SchedulerT, a scheduler for serverless functions. 
% \Scheduler has two key characteristics. 
% First, to avoid head-of-line blocking due to high function execution time variability, \Scheduler uses a combination of early binding and processor sharing scheduling at individual worker machines. 
% Second, \Scheduler uses a hybrid load balancing approach that improves consolidation at low load while employing least-loaded balancing at high load to retain high performance. 
% We implement \Scheduler for Apache OpenWhisk and demonstrate that, for the case of highly skewed workloads, characteristic of real-world traces, it achieves up to 85\% lower slowdown and 10\% higher throughput compared to OpenWhisk's default scheduling policy.

%!TEX root =asplos22-paper.tex  
%-------------------------------------------------------------------------------
\section{Introduction}
\label{sec:introduction}
%-------------------------------------------------------------------------------
Function-as-a-Service computing (FaaS) is becoming increasingly popular, primarily due to its ease of use~\cite{stateofserverless}.
Users just need to write functions in a high-level language, specify events and endpoints as execution triggers, and pay only for the resources used during function execution at fine (sub-second) granularity.
These functions are commonly called \textit{serverless functions} since infrastructure tasks such as resource provisioning, scheduling, scaling, and security are handled by specialized computing platforms~\cite{berkeley} without burdening the users.
These platforms exist both as part of cloud providers' offerings (AWS
Lambda~\cite{awslambda}, Google Cloud Functions~\cite{googlefunctions}, Azure
Functions~\cite{azurefunctions}) and as open-source frameworks that can be
deployed anywhere~\cite{openwhisk,kubeless,openfaas,openlambda,archipelago}.

The ease-of-use, cost-efficiency, and elasticity of serverless functions have led to the migration of many workloads to serverless platforms.
Serverless applications range from simple event processing functions~\cite{serverlessrepo} to complex frameworks that spawn a large number of tasks for workloads such as video processing~\cite{excamera, sprocket}, compilation~\cite{gg}, and machine learning~\cite{pywren,numpywren,infaas}.

Recent research has optimized various aspects of serverless platforms like the function startup-latency~\cite{catalyzer,seuss,firecracker,sock,msrserverless}, memory footprint~\cite{sock,firecracker,slacker}, and inter-function communication~\cite{serverlessclusters,coregranular,sand,ephemeralendpoints}.
In this paper, we focus on {\it scheduling serverless functions across and within machines}, an optimization critical to FaaS performance.
Our analysis is guided by a real-world production trace of Azure Functions~\cite{msrserverless}.
These real-world serverless functions are characterized by highly-variable
execution times, burstiness, and skewed invocations.
We take a principled approach and create {\it a taxonomy of scheduling policies}  that encompasses a broad set of techniques drawn from prior work on cluster and task scheduling and existing serverless frameworks.

Using simulation, we explore the policy space and conclude that commonly used techniques are sub-optimal for the characteristics of serverless functions:
First, we show that \emph{even idealized Late Binding is sub-optimal for
highly-variable workloads as short invocations can get stuck in the scheduler queue,
waiting for longer ones to finish execution}.
We can eliminate this problem if all invocations make forward progress using a processor-sharing policy.
To do so, we need to schedule invocations for execution as soon as they enter the system, i.e., use early binding.
This conclusion goes against the conventional wisdom that late binding should
always be preferred~\cite{pigeon,sparrow} or approximated~\cite{r2p2,sol}.
Second, \emph{we conclude that both random and locality-based load balancing across servers are
ineffective at high load; least-loaded balancing is necessary to avoid
performance loss due to load imbalance}.
Third, we identify the practical issues when using least-loaded balancing in real-world systems like Apache OpenWhisk~\cite{openwhisk}.
At low loads, the spreading of function invocations leads to the usage of more servers than necessary and more function cold starts. Hence, it achieves low hardware efficiency and increases function execution time.  

Based on these findings, {\it we develop \SchedulerT, a locality-aware hybrid scheduler that consolidates function invocations at a low load while reverting to least-loaded balancing at high load.}
%\neeraja{given what's in the previous para, I would rather say, "Based on these findings" instead of "to address these shortcomings"}
By consolidating invocations only at low load, \Scheduler achieves higher efficiency without sacrificing performance.
By reverting to least-loaded balancing for higher loads, it reduces queuing in overload situations.
\Scheduler is both load and locality aware; it takes into account locality when making scheduling decisions without sacrificing resource efficiency or causing load imbalance.
\Scheduler also incorporates the other lessons from our analysis by \textit{combining early binding with processor-sharing} to avoid head-of-line blocking.

We implement \Scheduler for Apache OpenWhisk~\cite{openwhisk}, a popular
open-source serverless platform.
We evaluate \SchedulerT's performance on workloads modeled after
Azure and Twitter production traces as well as workloads with characteristics similar
to emerging use cases of the serverless paradigm such as interactive analytics.
%\neeraja{next sentence is too long, can be broken into two.}
We focus our analysis on function slowdown, defined as function response time
divided by function execution time.
Slowdown is a performance metric suitable for characterizing the performance of workloads with highly-variable execution time~\cite{shinjuku,homa} and for comparing overheads and queuing delays due to
poor scheduling~\cite{sita}.
We show that \Scheduler achieves 85\% lower slowdown at low load - both at the median and the tail - than the currently-used OpenWhisk policy and supports 60\% higher load than commonly-used late binding approaches~\cite{pigeon,sparrow} for the Azure trace-based workload.
Being locality-aware, it achieves up to 50\% lower slowdown than a pure least-loaded policy while using up to 60\% fewer servers.
Additional experiments demonstrate that \Scheduler is robust to workloads with different function mixes and execution time distributions.
%\neeraja{you have to say explicitly which "other optimizations" you mean here. The sentence otherwise is assuming a lot of context you have in mind.}
%Furthermore, our scheduling improvements are largely orthogonal to other optimizations since they can be applied to different serverless platforms irrespective of the type of executors used, i.e.,  virtual machines, containers, or language runtimes.

The key contributions of this paper are the following:
\begin{itemize}[topsep=0pt]
    \item We introduce a taxonomy that decomposes scheduling approaches for serverless functions into three fundamental decisions and enables systematic research. 
    \item Using the characteristics of serverless functions as they appear in a
        real-world trace, we show that commonly used scheduling techniques,
        such as late binding and random load balancing, are sub-optimal for
        serverless workloads.
        %Contrary to existing belief, the scheduling
        %granularity and knowledge of function execution times are not important
        %factors.
    \item We use these conclusions to design \SchedulerT, a load and locality-aware scheduler that combines early binding, hybrid load balancing, and processor sharing at individual workers.
    \item We implement \Scheduler for Apache OpenWhisk~\cite{openwhisk}, the leading open-source serverless platform, and demonstrate performance and efficiency improvements over existing scheduling approaches.
%\neeraja{do we want to claim the "open-source"d-ness of Hermes?}
\end{itemize}

%\christos{drop if stressed for space} The rest of the paper is organized as follows.
%Section~\ref{sec:background} gives a brief introduction to serverless computing discussing its unique characteristics.
%In Section~\ref{sec:taxonomy} we introduce a scheduling taxonomy and use it to analyze the performance of different policies in simulation.
%Section~\ref{sec:design} discusses how we use the simulation insights to design and implement \Scheduler for OpenWhisk.
%Finally, in Section~\ref{sec:evaluation} we evaluate the performance of \Scheduler in a realistic OpenWhisk deployment.

%-------------------------------------------------------------------------------
%\section{An Introduction to Serverless Computing} 
\section{Background} 
\label{sec:background}
%-------------------------------------------------------------------------------
%\neeraja{Should we simply title this section as "background"?}
\subsection{Life-cycle of a Serverless Function}
\label{sec:background:lifetime}
\begin{figure}[h!]
\centering
\includegraphics[width=0.45\textwidth]{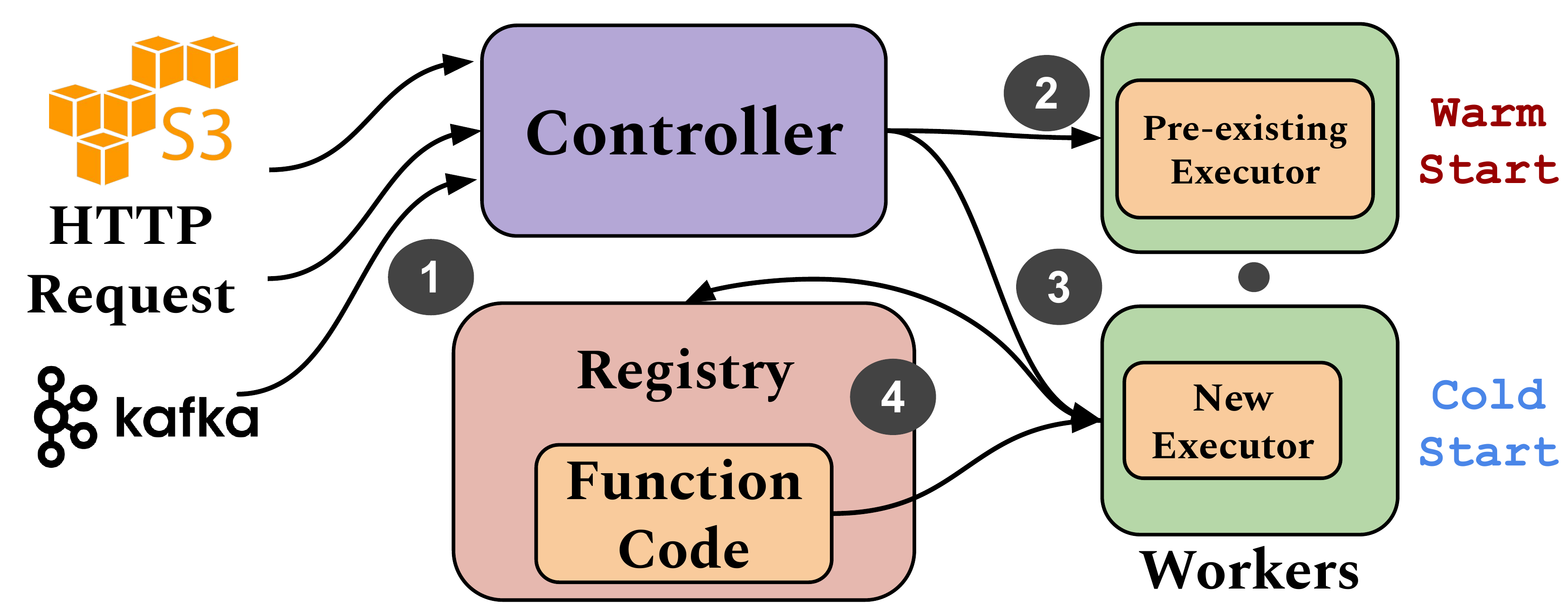}
\caption[width=0.49\textwidth]{Generic Serverless Platform Architecture
\label{fig:motivation:openwhisk}}
\vspace{-1.0em}
\end{figure}

%We use the setup of a generic serverless platform, shown in Figure~\ref{fig:motivation:openwhisk}, to describe the life-cycle of a serverless function invocation:
Figure~\ref{fig:motivation:openwhisk} describes the life-cycle of a serverless
function invocation in a generic serverless platform.

%\vspace{-1.0em}

\begin{itemize}
    \item Functions can be triggered in various ways, such as through HTTP requests, writes to a message queue, timer events, or uploads to cloud storage. Each function invocation is received by a Controller \ding{182} that terminates SSL connections and enforces access control and rate limiting.
    \item The Controller then schedules invocations to Workers for execution.
        One option is to forward the invocation to a Worker with a "warm"
        executor (VM~\cite{firecracker}, container~\cite{openwhisk}, or
        lightweight process~\cite{faasm}) running the function’s code
        \ding{183}. In this case, little initialization is needed and there is
        low overhead; this is a “warm start” of the serverless function.
    \item It is possible that there is no pre-existing executor for the invoked function or that all such executors are full. In that case, the Controller forwards the invocation to a Worker with available compute and memory capacity \ding{184}.
    \item When a Worker with no pre-existing executors receives an invocation,
        it fetches the function code from a registry, and starts a new executor
        \ding{185} incurring additional overhead. This is called a "cold start"
        of a function.
\end{itemize}

There can be many variations of this architecture.
In some platforms, there is a separate load balancing module that manages the scheduling of invocations to Workers.
It is also possible that some data store is used to persistently store critical
information, e.g., access control lists.
Persistent messages queues can be used instead of simple network protocols such as HTTP for Controller-Worker communication.

Regardless of the exact architecture, scheduling decisions are taken in two components, the \textit{Controller} and the \textit{Workers}.
The Controller decides how invocations are scheduled to Workers while individual Workers decide how their resources are distributed among the invocations they execute.

\subsection{Characteristics of Serverless Functions}
\label{sec:background:characteristics}
Microsoft recently released a  production trace of its serverless offering~\cite{msrserverless}.
The trace includes data about 445M function invocations across Microsoft Azure’s entire infrastructure over a 14-day period.
We use this trace along with information published by other cloud providers~\cite{stateofserverless} to infer the following characteristics of serverless functions:\\
\textbf{Short and Highly-variable Execution Times: } Functions typically live in the range of hundreds of milliseconds to minutes, and are billed at sub-second granularity.
The Azure trace showed that function execution times can be modeled using a heavy-tailed Log-normal distribution indicating extreme variability.
The median execution time is only 600 milliseconds while the 99\% execution time is more than 140 seconds.
Similar observations were made in AWS Lambda~\cite{stateofserverless}.\\
\textbf{Skewed Function Popularity: } Just 0.6\% of the functions account for 90\% of
the total invocations in the Azure trace.\\
\textbf{Burstiness: } Arrivals rates of individual functions are bursty with an average burstiness index of -0.26, close to that of a Poisson process~\cite{burstiness}. However, the total number of invocations across Microsoft's data centers does not vary much over time, following diurnal patterns that are characteristic of large-scale cloud systems~\cite{msrserverless,resourcecentral,borgnextgeneration}.
%\textbf{Statelessness: } Application state is stored remotely~\cite{pocket,ephemeral,shuffling} and hence locality does not matter other than avoiding the cold-start cost in case the function code needs to be fetched from a remote registry.

\subsection{Mismatch between Characteristics and Scheduling of Serverless Functions}
\label{sec:background:existing}
Most of the prior research on serverless platforms has focused on mechanisms that reduce the function cold-start cost and frequency.
Specialized virtual machines~\cite{firecracker}, lightweight and secure operating systems and language abstractions~\cite{faasm}, snapshots~\cite{vhive}, and check-pointing methods~\cite{catalyzer,seuss} have been able to reduce the start-up cost to less than 10 milliseconds while various techniques have been used to cache function data~\cite{faascache,ofc,faast} or predict future invocations~\cite{msrserverless} and avoid cold starts altogether.

Much less focus has been given to scheduling policies for serverless functions.
Scheduling has proven to be a critical factor to the performance of cloud systems~\cite{sparrow,pigeon,quasar,tarcil,monotasks,ray}.
Serverless platforms can also be considered as Remote Procedure Call (RPC) systems; better scheduling has provided up to an order of magnitude performance improvement for such systems~\cite{shinjuku,zygos, r2p2}.
Nevertheless, existing schedulers use techniques and policies that are ill-suited for the characteristics of serverless functions:\\
\textbf{Existing Schedulers for Serverless functions:}
Apache OpenWhisk~\cite{openwhisk} is one of the most widely used open-source serverless platforms. 
However, its default scheduler employs pure locality-based load balancing, co-locating invocations of the same function to a randomly-selected Worker without taking account of the load, a technique shown to be unable to handle \textit{highly-skewed workloads}~\cite{netcache, pegasus}.\\
\textbf{Kubernetes\-/based Frameworks:}
Many serverless platforms such as OpenFaaS~\cite{openfaas}, Kubeless~\cite{kubeless}, and vHive~\cite{vhive} are built on top of Kubernetes and employ a different type of scheduling.
They treat serverless functions similarly to classic server workloads and employ auto-scaling.
Once the CPU or memory utilization exceeds some pre-defined threshold, more workers are spun up to handle the extra load.
If utilization drops below a threshold, the load is consolidated to fewer workers.
Such coarse-grain and reactive policies cannot handle the \textit{burstiness} associated with serverless workloads, and lead to high tail latency and slowdown~\cite{shenango,ixcp}.\\
\textbf{General Task Schedulers:}
Tasks scheduled by scientific computing and analytics frameworks have common characteristics with serverless functions, i.e., short execution times and burstiness.
Some of these frameworks focus on improving data locality by queuing tasks or migrating them between machines so that they execute close to their corresponding data~\cite{canary,ray}.
However, the overhead associated with cold-starts~\cite{peeking} makes migrations unappealing for serverless functions.
Task schedulers that avoid task migrations or do not place hard locality constraints could prove more suitable. 
Sparrow~\cite{sparrow} and Pigeon~\cite{pigeon} - two popular schedulers for
data analytics frameworks - employ late binding, i.e., task-to-worker
assignments are delayed until workers are ready to run the task.
%with the goal of minimizing queuing.
However, late binding is sub-optimal for workloads with \textit{highly-variable} execution times: shorter tasks can suffer slowdown due to getting stuck behind tasks with much longer execution times.
Interleaving of shorter and longer tasks is desirable for such workloads to
avoid this phenomenon, known as head-of-line blocking~\cite{homa,shinjuku}.
Hawk~\cite{hawk} and Eagle~\cite{eagle} improve upon Sparrow by differentiating between short and long tasks.
A set of nodes is set aside for short tasks to avoid head-of-line blocking.
Long tasks are centrally scheduled while short tasks use distributed schedulers and the probing techniques employed by Sparrow.
Scheduling in these systems requires (i) knowledge of each task's execution time and (ii) work-stealing to compensate for low-quality scheduling decisions made by distributed schedulers.
These requirements make such schedulers incompatible with serverless functions due to their irregular execution times and the high cold-start cost.\\
\textbf{Cluster Schedulers:}
There is a large body of work on cluster schedulers that can manage thousands of jobs, belonging to different users and applications.
Some of these schedulers (Borg~\cite{borg}, Yarn~\cite{yarn}, Mesos~\cite{mesos}, Mercury~\cite{mercury}) allow different frameworks to request and receive cluster resources, leaving the allocation of resources to tasks to the individual framework.
Such resource managers can be used by serverless frameworks but they are orthogonal to serverless function scheduling.
The serverless framework will still need to decide when to request resources, i.e., servers or CPU cores, and how to schedule function invocations to these resources.

Other cluster schedulers, despite being more suitable for scheduling individual tasks, focus more on long-running tasks incurring too high overhead to short-running serverless functions.
The Quincy~\cite{quincy} and Firmament~\cite{firmament} frameworks model scheduling as a min-cost max-flow (MCMF) optimization problem over a flow network.
While these frameworks provide high-quality decisions, they suffer from long scheduling delays and often require job migrations exacerbating the cold-start problem of serverless functions.
Stratus~\cite{stratus} tightly packs tasks to machines to improve efficiency and reduce costs.
We use a similar approach in one of the two modes of operation of Hermes.
Medea~\cite{medea} optimizes the scheduling of long-running applications, an important objective which is however irrelevant to short serverless function invocations.

\section{Taxonomy of Scheduling Policies}
\label{sec:taxonomy}

\subsection{Scheduling Analysis}
\label{sec:taxonomy:analysis}
The first step to designing a scheduler for serverless functions is to identify the scheduling decisions made during the life-cycle of a function invocation (\S~\ref{sec:background:lifetime}):
\begin{enumerate}
    \item When should an invocation be scheduled to a Worker?
    %For example, if there is no available container for a specific function, should the scheduler wait until one becomes available, or should it schedule the invocation to run on a different worker, paying for the cold start cost? 
    %\item At what granularity should scheduling take place? Per-core or per-server?
    \item Which Worker should handle each invocation?
    \item Which intra-Worker scheduling policy should be used?
\end{enumerate}
The first two decisions are enforced at the \textit{Controller} while the third is enforced at individual \textit{Workers}.

We now introduce a taxonomy that describes the policies resulting from different answers to the aforementioned scheduling questions.
We use this taxonomy to formalize our analysis in the following sections and identify the policies that cater to the characteristics of serverless functions.\\
\textbf{When should an invocation be scheduled to a Worker?} There are two
major options, early and late binding.
In early binding scheduling, no queuing takes place at the Controller, i.e., function invocations are assigned to Workers for execution as soon as they reach the Controller.
    Early binding is commonly used when queuing at the Controller would consume too much memory to store the functions' arguments, such as in OpenWhisk~\cite{openwhisk}, when the Controller-Worker communication latency is high and needs to be hidden (Canary~\cite{canary}, R2P2~\cite{r2p2}, Mind the Gap~\cite{mindthegap}), or in some distributed settings where schedulers do not have a global view of the system and must choose the best available among a limited number of nodes (Hawk~\cite{hawk}, Eagle~\cite{eagle}).
If late binding is used, invocations are queued at the Controller until there are available resources to guarantee their immediate execution.
Late binding is assumed to provide better performance than early binding since it completely avoids load imbalances at Workers and thus is preferred in systems that can tolerate the overheads associated with it (Firmament~\cite{firmament}, Quincy~\cite{quincy}, Medea~\cite{medea}, Sparrow~\cite{sparrow}, Pigeon~\cite{pigeon}).\\
%\textbf{What scheduling granularity should be used? } Another dimension is the granularity at which scheduling is done.
%Most existing systems~\cite{openwhisk, canary, sparrow, ray, pigeon, hawk, eagle, kairos, firmament, quincy} schedule at server granularity; invocations are assigned to individual virtual or physical machines.
%Recently, the use of core-granular scheduling~\cite{coregranular, r2p2, monotasks} was proposed to reduce interference and give the Controller finer-scale control.\\
\textbf{Where should a function invocation execute?}
Regardless of whether early or late binding is used, we need to determine a load balancing policy that selects a Worker to run a function on.
This is the decision most cluster schedulers (Quincy~\cite{quincy}, Firmament~\cite{firmament}, Medea~\cite{medea}, Stratus~\cite{stratus}) make.
One approach - used by OpenWhisk~\cite{openwhisk} - is to optimize for locality, packing invocations of the same function in as few Workers as possible to minimize the number of cold starts.
Another approach that provides good load balancing in expectation with low overhead is random assignment~\cite{ix}.
Finally, there is a group of load-aware policies, e.g., always selecting the least-loaded Worker.
Both large-scale task schedulers (Hawk~\cite{hawk}, Eagle~\cite{eagle}, Kairos~\cite{kairos}) and small-scale microsecond-level request schedulers (RackSched~\cite{racksched}, R2P2~\cite{r2p2}) have used or approximated least-loaded balancing.\\
\textbf{Which intra-Worker scheduling policy should be used? } If early binding
is used, there is queuing at the Workers and a scheduling policy must control the way each Worker's resources are used by different invocations.
Different policies are optimal for different types of workloads.
    The Shortest-Remaining-Processing-Time (SRPT) scheduling policy has long been known to be optimal for minimizing mean response time and is commonly used in environments where information about the task execution time is available (Eagle~\cite{eagle}).
It is hard to use SRPT in most setups - including serverless function scheduling - because the processing time is rarely known a priori.
    Hence, some schedulers, e.g., Kairos~\cite{kairos}, attempt to approximate it by using the Least Attained Service scheduling policy~\cite{fbqueue}.
A policy with similar characteristics is Processor Sharing (PS) where each task receives an equal share of the processor's capacity.
Linux's Completely Fair Scheduler (CFS)~\cite{cfs} is an approximation of PS that is applicable to real systems.
Since most systems that employ early binding delegate scheduling to the operating system, they end up using CFS - and thus approximate PS - for intra-Worker scheduling.
    Another practical policy is First-Come-First-Serve (FCFS) which has been used by both task schedulers such as Hawk~\cite{hawk} and low-latency systems (R2P2~\cite{r2p2}, HovercRaft~\cite{hovercraft}, IX~\cite{ix}) due to its simplicity and low overhead.

\paragraph{Notation: }We use notation similar to Kendall’s from standard queuing theory~\cite{kendall} to describe the different scheduling policies that can be created by combining the aforementioned options.
We describe a policy through the use of 3 parameters, $T / LB / S$, where:
\begin{itemize}
    \item \textbf{T} describes whether early or late binding is used. The two possible values for $T$ are $E$ and $L$. 
    $E$ denotes early binding and $L$ denotes late binding. 
    %\item \textbf{G} describes the scheduling granularity. The two possible values for $G$ are $S$ and $C$. 
    %$S$ denotes server-granular and $C$ denotes core-granular scheduling. 
    \item \textbf{LB} describes the load balancing policy used to select an
        available Worker. Three possible values for $LB$ are $LOC$ (locality-based balancing), $LL$ (least-loaded balancing) and $R$ (random balancing).
    \item \textbf{S} describes the scheduling policy used in the Workers. The two practical such policies that we use are Processor-Sharing (PS) and First-Come-First-Serve (FCFS). We discuss policies that require knowledge of the request execution time in \S~\ref{sec:simulation:srpt}.
\end{itemize}
For example, with this notation we can describe the OpenWhisk scheduler that
uses early binding ($E$), locality-based load balancing
($LOC$), and processor sharing at the Worker servers ($PS$) as $E/LOC/PS$.

The mismatch between existing schedulers and the needs of serverless
functions as demonstrated in the Microsoft Azure trace (\S~\ref{sec:background:existing}) prompts us to go back to the drawing table. 
Guided by the taxonomy, we use simulation to explore the scheduling policy space and draw conclusions that help us design an effective policy for serverless function scheduling.

\subsection{Simulation Setup}
\label{sec:simulation:setup}
We built a discrete event simulator that allows us to rapidly evaluate and compare different scheduling policies and architectures.
Serverless function invocations are generated following tunable distributions for the inter-arrival and the execution times.
We focus our analysis on the Azure trace and thus use highly-variable Log-normal execution times and open-loop Poisson arrivals~\cite{treadmill, openvsclosed, msrserverless}.
A Controller schedules incoming activations to Worker servers.
Each Worker has a number of cores determining the number of invocations it can execute in parallel and a memory capacity that limits the maximum number of invocations - running and waiting - it can host at any point in time.
To mimic our testbed, we set the capacity of each server in terms of invocations to be 8$\times$ its number of cores.
We implemented different load balancing and intra-Worker scheduling policies.
The goal of the simulator is to help us understand the behavior of serverless workloads, not to perfectly mimic the behavior of a real system.
To model the trace's skew, invocations belong to 50 distinct functions, one contributing 98\% of the load and the rest making up the rest of the load equally.
The simulator does not model overheads such as the container start-up time; in Section~\ref{sec:evaluation} we show that the simulation results hold in a real deployment where these overheads are considered.
%We plan to open-source the simulator after the publication of this paper.

\subsection{Pruning the Policy Space}
\label{sec:simulation:single}
We start our analysis from a simple example where we have \textbf{four
Worker servers with $12$ cores each}.
We compare all 12 possible policies we defined in Section~\ref{sec:taxonomy:analysis}:
\begin{itemize}
    \item \textbf{$L/*/*$~(Late Binding):} Using late binding scheduling,
        function invocations are scheduled to a Worker in an FCFS manner only
        when resources are available. If no resources (cores) are available,
        invocations are queued at the Controller. The simulator does not
        capture second-order effects such as interference. Hence, the
        the load balancing and the intra-Worker
        scheduling policy do not matter when late binding is used; if a
        function invocation is scheduled to a Worker, our simulator assumes that it will run uninterruptedly. This policy is used by popular task scheduling systems such as Pigeon~\cite{pigeon} and Sparrow~\cite{sparrow}.
    \item \textbf{$E/LL/FCFS$~(Early Binding / Least-Loaded Balancing / FCFS):} The Controller implements a Join-Shortest-Queue policy where invocations are queued locally at each Worker where they execute FCFS. This policy is used by some RPC systems such as R2P2~\cite{r2p2}.
    \item \textbf{$E/LL/PS$~(Early Binding / Least-Loaded Balancing / PS):} Similar to the previous policy, but the queued invocations timeshare each Worker.
    \item \textbf{$E/LOC/FCFS$~(Early Binding / Locality-Based Balancing / FCFS):} The Controller forwards all invocations belonging to the same function to a randomly-selected Worker where they execute FCFS. If the Worker's capacity is reached, we select a different random Worker.
    \item \textbf{$E/LOC/PS$~(Early Binding / Locality-Based Balancing / PS):} Similar to the previous policy, but the queued invocations timeshare each server.
    \item \textbf{$E/R/FCFS$~(Early Binding / Random Load Balancing / FCFS):} The  Controller randomly assigns invocations to Workers where they execute FCFS.%This policy is used by some intra-server dataplane operating systems~\cite{ix} that randomly assign incoming network flows to cores for execution.
    \item \textbf{$E/R/PS$~(Early Binding / Random Load Balancing / PS):} Similar to the previous policy, but the queued invocations timeshare each core.
\end{itemize}

\begin{figure}[h!]
    \centering
    \subfloat[99\% Latency]{
        \includegraphics[width=\linewidth]{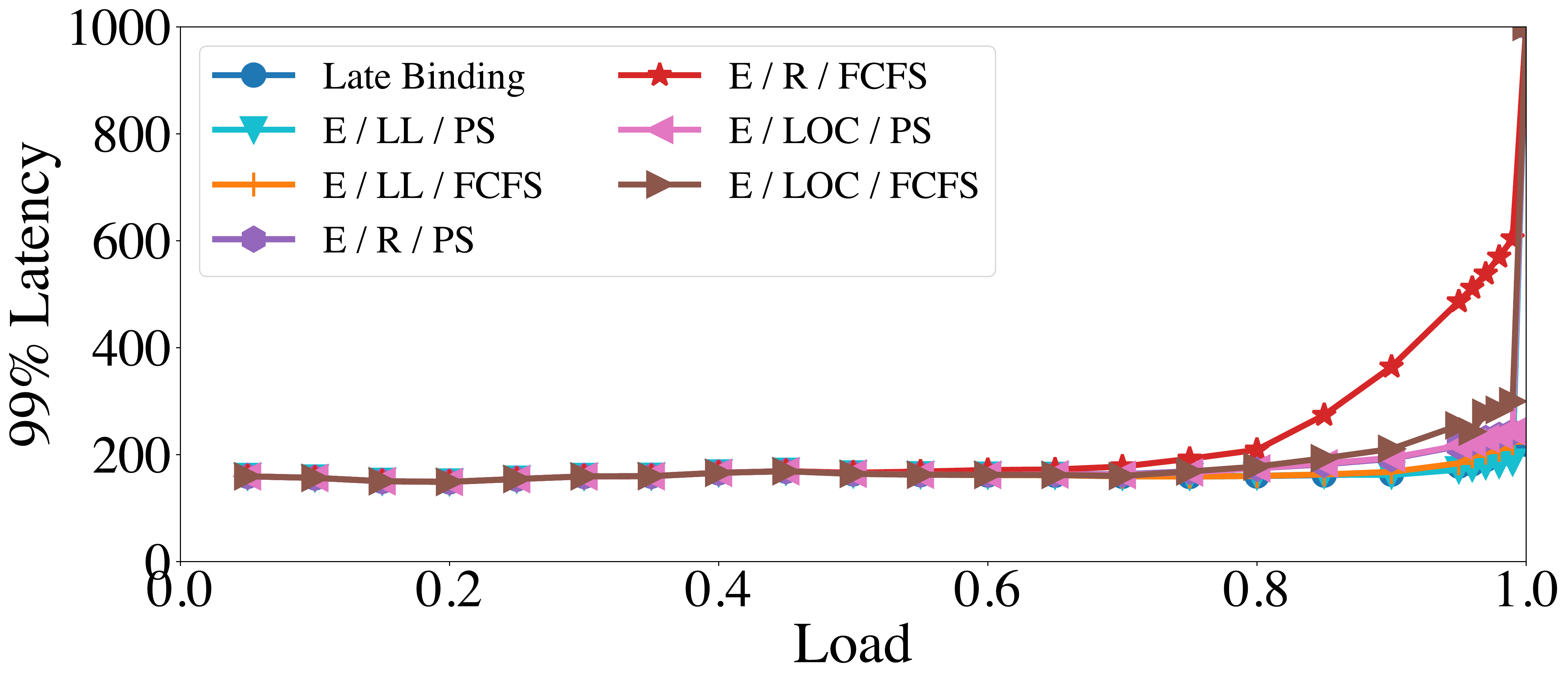}
    }\\
    \vspace{-1.0em}
    \subfloat[99\% Slowdown]{
        \includegraphics[width=\linewidth]{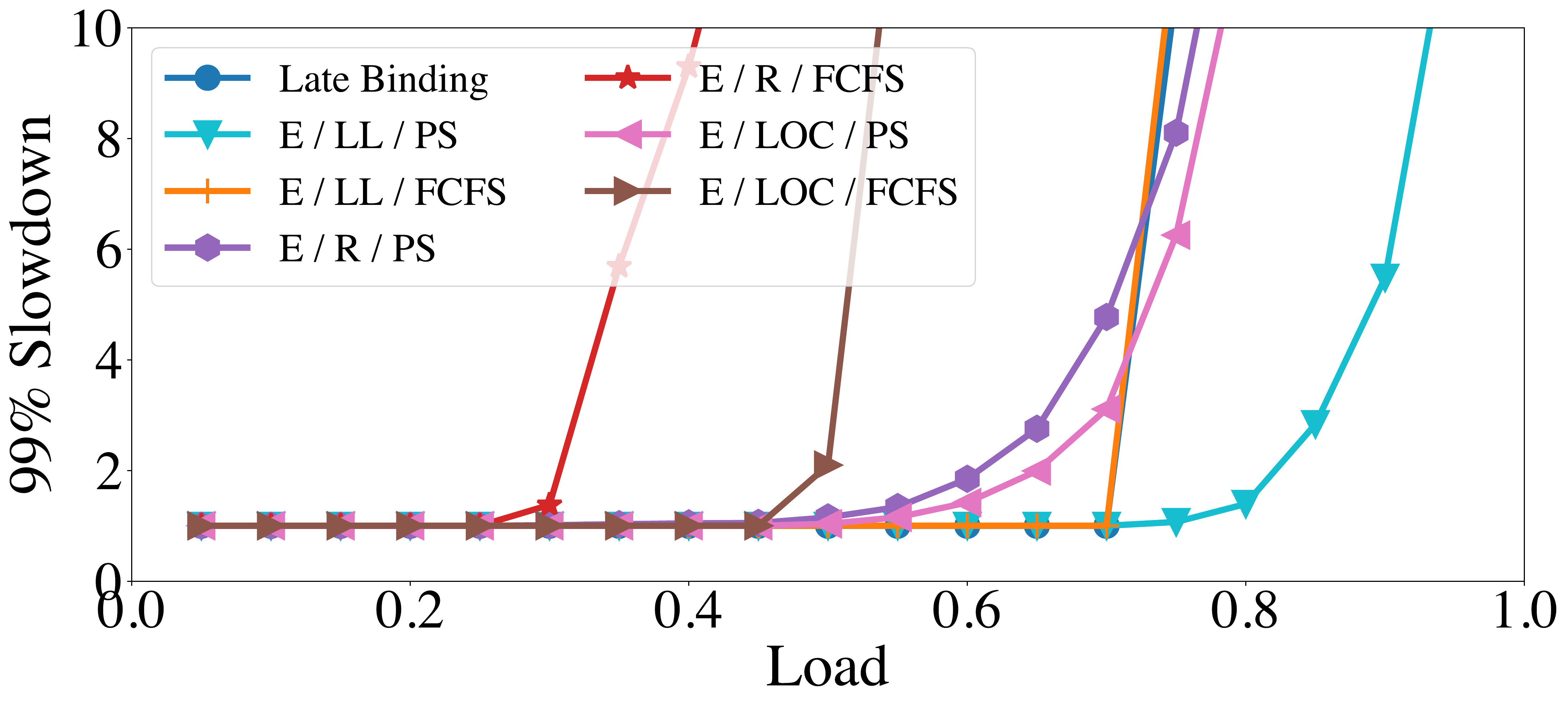}
    }
    \caption{Tail latency and slowdown as a function of the load for a synthetic
    workload with the characteristics of the Azure trace~\cite{msrserverless}, i.e., Log-normal execution times with parameters $\mu = -0.38$ and $\sigma = 2.36$, in a single server.}
\label{fig:scheduling:basic}
\end{figure}

In Figure~\ref{fig:scheduling:basic}a we can see the 99\% latency for the different policies as a function of the offered load indicated by the fraction of the server's capacity. 
For a load greater than 1, the system becomes unstable and the queues expand indefinitely. 
We observe that most policies perform similarly and almost optimally; the tail latency explodes for a load close to 1.
%$E/C/R/PS$ uses random load balancing which allows for transient imbalances affecting the tail latency.
%Similarly, $E/C/LOC/PS$ leads to increased tail latency as some cores are overloaded due to the high skew.
%While $E/C/LL/FCFS$ keeps the load perfectly balanced, it is possible that short invocations get stuck behind long ones at individual cores leading to worse tail latency due to the head-of-line blocking effect.
%Finally, $E/C/R/FCFS$ and $E/C/LOC/FCFS$ have poor performance, combining load imbalances and head-of-line blocking due to per-core FCFS scheduling.
%Similar behavior has been observed in data-plane operating systems with per-core queues~\cite{shinjuku, zygos, mica}.
Based on these latency results, we get the false impression that scheduling does not matter much.
%Late Binding, as used by systems like Sparrow~\cite{sparrow} and Pigeon~\cite{pigeon}, performs well for serverless workloads.

Next, we study the 99\% slowdown for the same setup.
Slowdown is defined as:
\vspace{-0.5em}
$$\text{slowdown} = \frac{\text{function latency}}{\text{function execution time}}$$
\vspace{-0.5em}

The closer this ratio is to $1$, the smaller the delay a user experiences compared to what she expected based on the function execution time.
Slowdown is a metric commonly used in the networking literature when reasoning
about the performance of request-based applications~\cite{homa,shinjuku}.
It is more insightful than latency as a performance metric as it can reveal pathological cases of system behavior.
For example, in a workload where half of the requests are short, e.g., 100ms, and half are long, e.g., 10sec, the 99\% latency - a metric commonly used for performance analysis - is not useful.
If the 99\% latency is reported to be 10sec, that would give users no information about the performance of their short requests.
However, if the 99\% slowdown is reported to be 1.5, users would know that most
of both the longer and the shorter requests do not experience excessive
delays.\\

We observe that some policies, e.g., Late Binding and $E/LL/FCFS$, perform significantly worse in terms of slowdown, while tail latency had painted a completely different picture.
If we used tail latency as the metric, we would assume everything is good and would not analyze further to find out the real problem.
Hence, in the rest of this paper we use slowdown as the main metric for our analysis.
We make the following observations in Figure~\ref{fig:scheduling:basic}b:\\
\textbf{Observation 1: } Processor Sharing in the Workers significantly outperforms FCFS-based scheduling - even Late Binding - regardless of the load-balancing policy used.
The optimality of PS over FCFS in terms of tail performance for heavy-tailed workloads has been proven by Boxma and Zwart~\cite{tailinscheduling}.
Intuitively, functions with light-tailed execution time take about the same time to finish.
Therefore, one can minimize the slowdown by running invocations to completion in their arrival order~\cite{ix}.
In the case of heavy-tailed workloads, the slowdown can increase dramatically if short invocations get stuck behind longer ones.
PS allows short invocations to receive a slice of the processor and bypass the longer ones.
However, PS is incompatible with late binding.
Invocations need to be assigned to servers/cores for execution as soon as they enter the system.
    Otherwise, they are not able to time-share compute resources and suffer from head-of-line blocking.\\
\textbf{Lesson Learned 1: } \textit{Due to head-of-line blocking, Late Binding schedulers and Early Binding schedulers that use FCFS cannot handle the high execution time variability present in serverless workloads.}\\
\textbf{Observation 2:} The slowdown for both random and locality-based load balancing starts to go up for a load as low as 0.55.
Random load balancing results in high latency and slowdown for invocations with highly-variable execution time since it is likely for imbalances across Workers to appear.
Similarly, locality-based balancing - as used by Openwhisk - leads to significant load imbalances.
Workers serving "hot" functions quickly become overloaded, leading to high slowdown even for a medium load.
Least-loaded balancing performs very well keeping the slowdown low for a load as high as 0.8.\\
\textbf{Lesson Learned 2: } \textit{Random and locality-based balancing, similar to what OpenWhisk is using, is ineffective at high load. Least-loaded balancing provides better performance.}

\subsection{Simple Scheduling can be Better}
\label{sec:simulation:srpt}
The two intra-Worker scheduling policies we considered so far,
Processor-Sharing and First-Come-First-Serve, do not assume knowledge of the
function execution times.
Even though this characteristic makes these policies usable in
any environment, one could argue that an intelligent scheduler could leverage
knowledge from past function invocations to predict future function execution
times, similar to what Shahrad et al.~\cite{msrserverless} did for
inter-arrival times.
To see if such knowledge benefits a serverless scheduler, we
evaluate the performance of a scheduler that assumes perfect knowledge of
function execution times and uses the Shortest-Remaining-Processing-Time
(SRPT) policy for intra-Worker scheduling.
We chose SRPT as it has been shown by Bansal and Balter~\cite{srpt} that it
performs better than PS in terms of mean and median slowdown.

\begin{figure}[h!]
    \centering
   \subfloat[]{
        \includegraphics[width=0.49\linewidth]{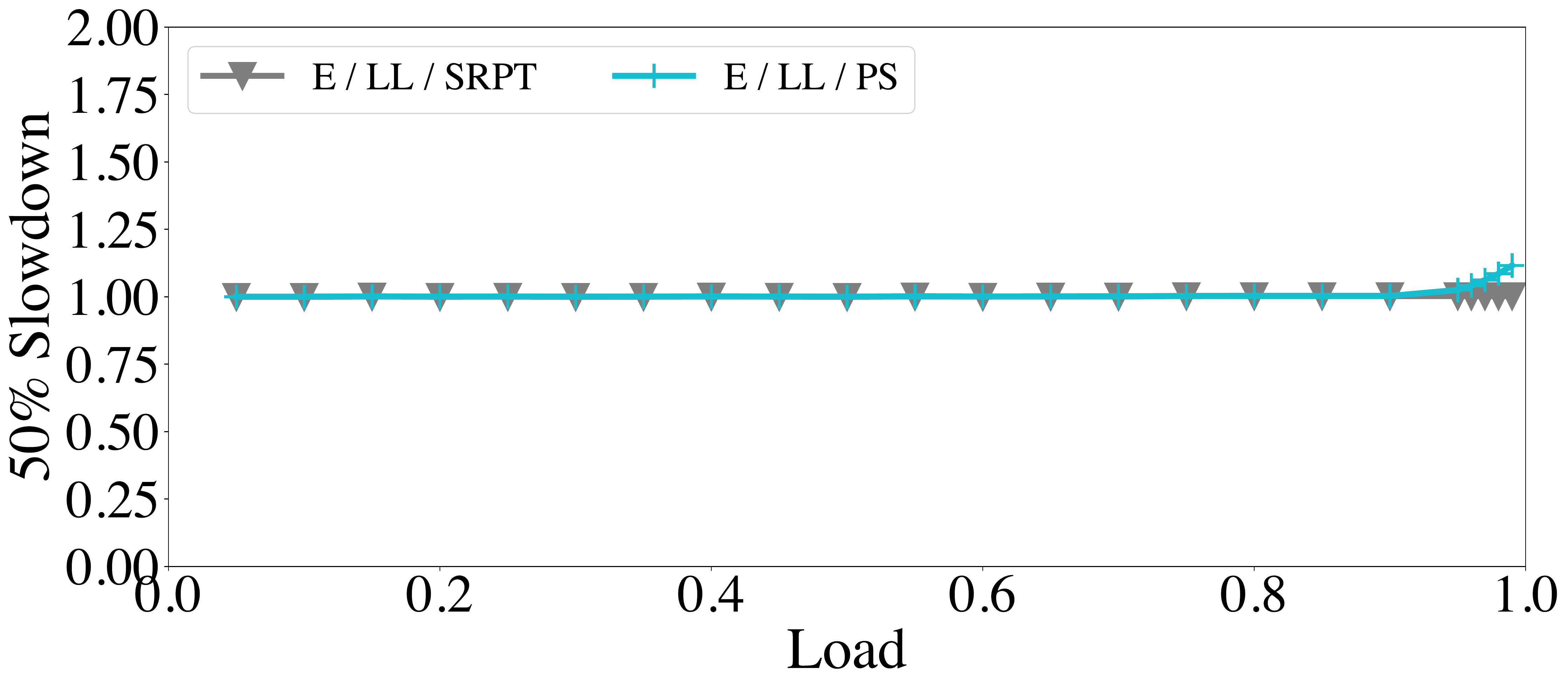}
    }
    \subfloat[]{
        \includegraphics[width=0.49\linewidth]{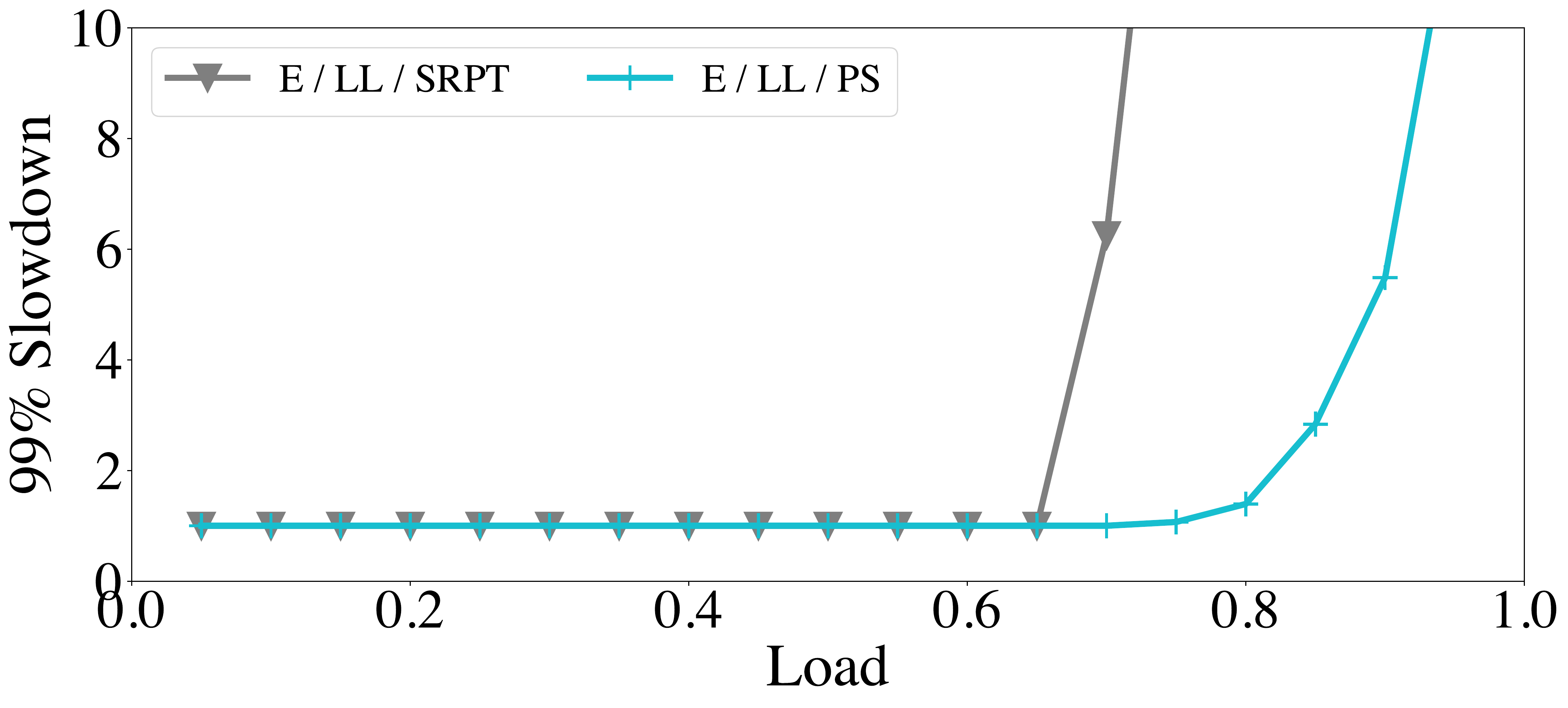}
    }
    \caption{Slowdown in a 4-server, 12-core setting under the SRPT and PS scheduling policies.}
\label{fig:scheduling:srpt}
\end{figure}

\textbf{Observation 3: }
In Figure~\ref{fig:scheduling:srpt}, we observe that, as expected,
the best performing SRPT-based policy ($E/LL/SRPT$) slightly outperforms the
best PS-based policy ($E/LL/PS$) in terms of median slowdown at high load.
We omit the worse-performing SRPT configurations for brevity.
However, we also observe that the SRPT policy performs significantly worse
than the PS-based one in terms of \emph{tail} slowdown (99\%) which is our metric of
interest.
The slowdown starts increasing for load as low as 0.7 for the $E/LL/SRPT$
scheduler while for the $E/LL/PS$ scheduler the slowdown is less than 10 for
load as high as 0.9.
While SRPT is optimal in terms of slowdown on average, its tail performance is
worse than that of PS at high load because it can lead to starvation for longer
requests.
We expect SRPT's performance to be worse in a realistic environment where
perfect knowledge of execution times is unavailable and scheduling decisions
depend on estimates.
\\
\textbf{Lesson Learned 3: }
We demonstrated that, counterintuitively, using perfect knowledge of the
execution time of each individual function invocation through SRPT, a
well-studied policy, does not offer performance benefits over PS
for \emph{tail} slowdown.
Hence, in the rest of the paper we consider only execution-time-agnostic
policies which are more generally applicable.

\subsection{Scheduling at Scale}
\begin{figure}[h!]
\centering
\includegraphics[width=0.45\textwidth]{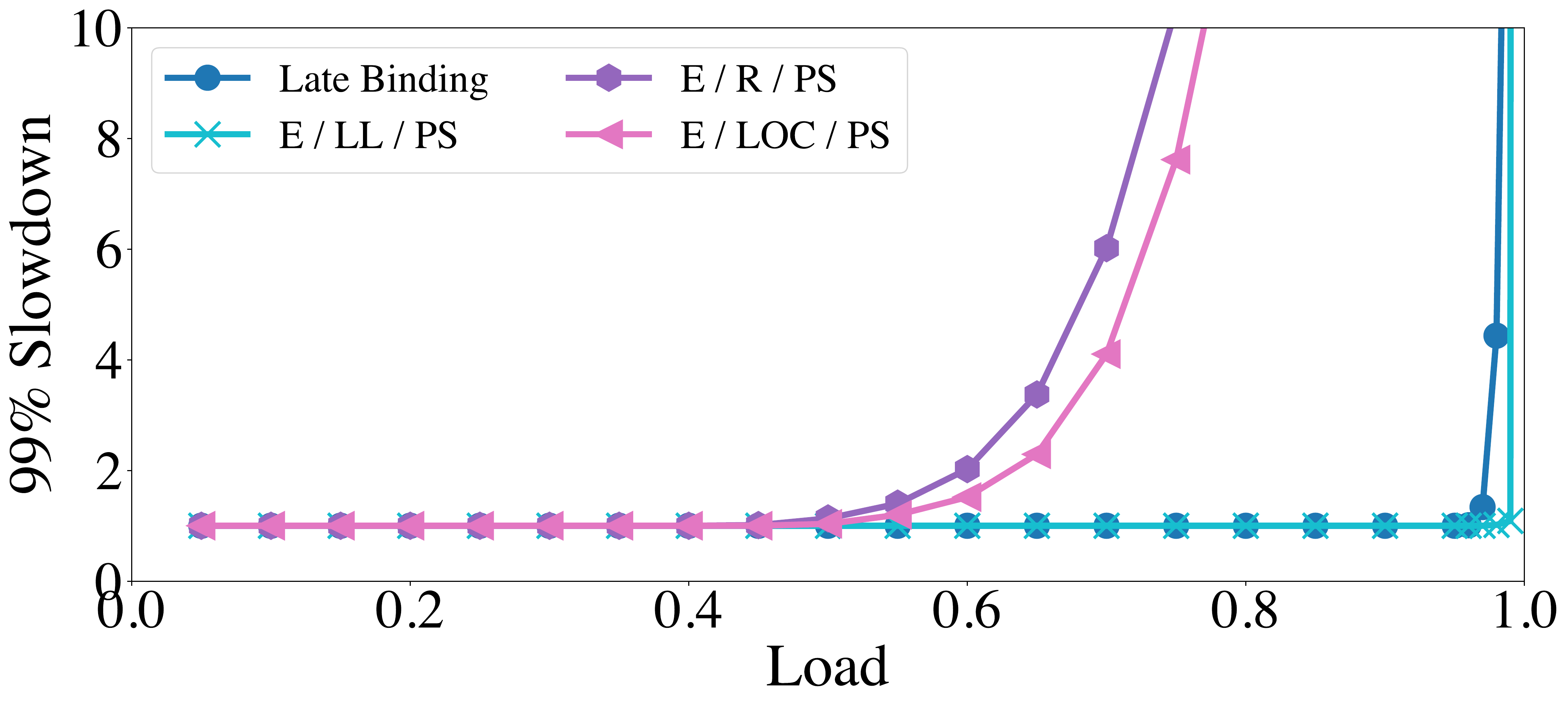}
    \caption[width=0.49\textwidth]{99\% slowdown in a 100-server, 12-core setting.
    \label{fig:scheduling:100workers_slowdown}}
\end{figure}

As Figure~\ref{fig:scheduling:100workers_slowdown} shows, our main conclusions are still valid for a large-scale, 100-server, 1200-core setup.
$E/R/PS$'s and $E/LOC/PS$'s slowdown explodes at a relatively low 0.6 load.
Late Binding performs much better than in smaller setups because occurrences of head-of-line blocking reduce as the number of Workers increases.
However, we see that $E/LL/PS$ still outperforms Late Binding at very high loads (>0.96).
That, together with the inherent complexities and overheads associated with Late Binding, convinces us that $E/LL/PS$ is the best policy to use regardless of the scale of the setup.

%-------------------------------------------------------------------------------
\section{Hermes Design}
\label{sec:design}
Based on the conclusions of the simulations in \S~\ref{sec:taxonomy}, we developed \SchedulerT, a scheduler that is both \textit{locality and load-aware}.
\Scheduler caters to the unique characteristics of serverless functions by combining \textit{early binding} and \textit{processor sharing} with \textit{hybrid load balancing} that adapts to changes in the load.
Moreover, it keeps track of warm available containers for each function, avoiding cold-starts when possible.
%, i.e., it adapts the load
%balancing policy it uses based on the load.
%-------------------------------------------------------------------------------%
\subsection{Why Hybrid Load Balancing is Necessary}
%According to the Azure trace, serverless functions are characterized by high execution time variability.
Based on the simulations, the ideal scheduling policy for a serverless workload combines \textbf{early binding} with \textbf{least-loaded} balancing and \textbf{processor-sharing} scheduling at the Workers ($E/LL/PS$). However, an $E/LL/PS$ policy suffers from practical shortcomings: % that can hinder its effectiveness for a real serverless platform.\\

\noindent\textbf{Low Resource Efficiency: } 
Least-loaded balancing spreads function invocations across all servers in a deployment. 
Particularly at low load, this results in poorly-utilized servers increasing costs~\cite{stratus}, despite existing strategies to harvest unused cloud resources.
Consolidating the workload in fewer servers is especially important in public clouds since performance metrics are hidden from the providers prohibiting them from easily harvesting spare resources without violating customer SLOs~\cite{scavenger,perfiso,harvest,pado,deflation}.

\noindent\textbf{Increased Cold Starts:}
Function code locality is an essential consideration for serverless schedulers since cold starts can cause high function invocation latency.
However, by spreading invocations across a larger number of servers, a scheduler that uses least-loaded balancing causes more function cold starts.
At low loads, there is no tendency for function invocations to be scheduled to a Worker with warm executors.

%The alternative is to use \textbf{late binding} (L/*/*). However, late binding approaches suffer from the inverse problem. 
%At low load, by scheduling function invocations to Workers until they reach capacity, they are able to consolidate load in a smaller number of servers without having to sacrifice slowdown.
%This allows them to achieve high efficiency.
%However, at high load, invocations are queued at the Controller waiting for compute resources to become available.
%This, together with the highly-variable execution times that are characteristic of serverless functions, can cause high slowdown. 

\begin{figure}[h!]
\centering
\includegraphics[width=0.45\textwidth]{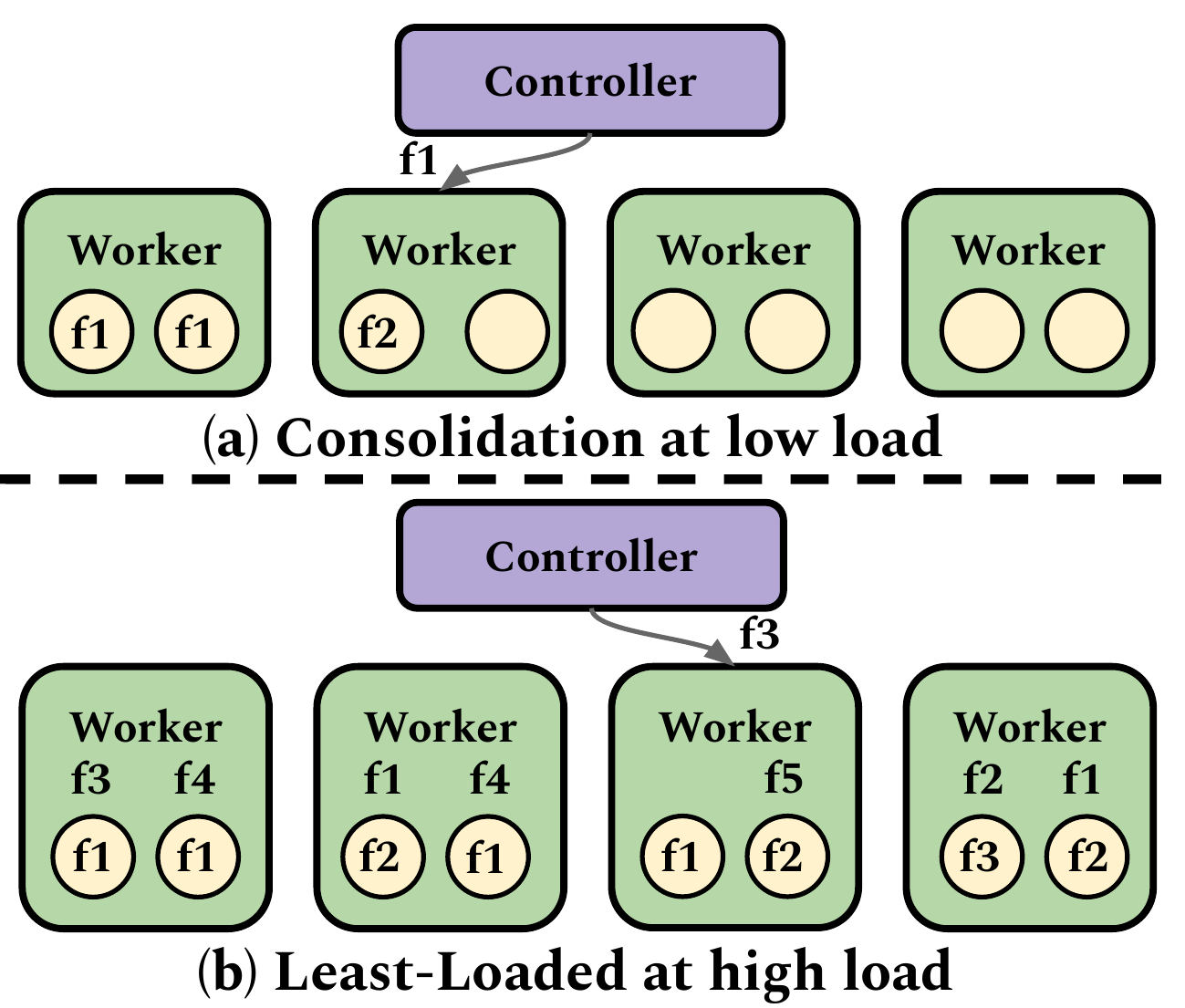}
\caption[width=0.49\textwidth]{\Scheduler (a) packs function invocations to machines in a core-aware manner at low load and (b) uses least-loaded balancing when all cores are full. The yellow circles represent each Worker's CPU cores.}
\label{fig:design:hybrid}
\end{figure}
\vspace{-2.0em}

\subsection{\Scheduler ($E/H/PS$)}
\Scheduler uses \textit{early binding and processor-sharing scheduling} to
avoid head-of-line blocking that can cause high slowdown at high load
\textit{without assuming knowledge of the function execution times}.
To address the shortcomings of least-loaded and late-binding schedulers, \Scheduler employs locality-aware \textit{hybrid load balancing} ($H$), using one of two different load balancing modes depending on the load.\\

\textbf{Low load:}
At low load, i.e., when there are available cores in some Worker, \Scheduler packs functions to Workers with available cores (Figure~\ref{fig:design:hybrid}a).
Starting from an arbitrary Worker, \Scheduler fills it with a number of invocations equal to the number of Worker cores.
As more invocations arrive, the same process is applied iteratively for all Workers in the deployment.
When ongoing invocations finish execution, \Scheduler directs new incoming invocations to the corresponding servers filling them up.
%If an invocation finishes execution in some previously full Worker, \Scheduler will direct incoming invocations there.
This policy of assigning up to $N$ invocations to each $N$-core Worker,
guarantees that there is no queuing at low load~\cite{sparrow} while it
consolidates function invocations.\\

\textbf{High load:}
When all Workers are filled with invocations equal to their number of cores, load balancing reverts to least-loaded (Figure~\ref{fig:design:hybrid}b).
This way, \Scheduler reduces queuing in overload situations.
\Scheduler can accurately make the distinction between the two modes of operation by synchronously communicating with each Worker.
This incurs low overhead since the Controller-Worker communication latency (O(1msec)) is much lower than function execution times (O((1sec)). \\
%Hence, it is safe to assume that the Controller synchronously knows each Invoker's load when making a scheduling decision.\\

\textbf{Locality:}
The hybrid load balancing described above alleviates the cold-start problem by consolidating invocations to a smaller number of servers.
To improve even further, we explicitly make \SchedulerT's load balancing policy locality-aware.
\Scheduler keeps track of the available warm containers in each Worker.
It first looks for a warm container in one of the non-empty Workers with available capacity when it operates at low-load mode.
If such a Worker exists, \Scheduler steers the invocation there.
If not, it prioritizes consolidation by selecting a non-empty Worker without a warm container instead of an empty Worker with a warm container.
When it operates under high load, \Scheduler breaks ties using locality.
If multiple Workers are equally loaded, it selects the one with a warm container for the incoming invocation if such a Worker exists.

In \S~\ref{sec:evaluation:coldstart} we show that \Scheduler improves locality compared to existing schedulers.
Moreover, \Scheduler is compatible with recently proposed predictive approaches~\cite{msrserverless} that reduce function cold starts.
In such designs, the controller spins up warm containers in anticipation of future function invocations.
The only difference for \Scheduler is that its hybrid load balancing will run earlier, i.e., when
the container is spun up instead of when the invocation arrives.\\

In summary, \Scheduler achieves the best of both worlds: high consolidation with low slowdown and low cold-start rate at low load while also maintaining low slowdown at high load.% (\S~\ref{sec:evaluation}).

\subsection{Scalability}
In order for cloud schedulers to scale, they need to be able to operate in a
distributed fashion.
\SchedulerT's design is compatible with common techniques
used in distributed schedulers, such as sharding and power-of-k choices.
In a sharded setup, each instance of \Scheduler can load-balance requests among
the subset of Workers it owns.
In a power-of-k choices architecture~\cite{mitzenmacher, sparrow}, when an
invocation reaches one of many scheduler instances, the scheduler samples k
Workers and chooses one of them to send the invocation to according to some
scheduling policy. \Scheduler can operate by using hybrid load balancing to
choose one of the k sampled Workers.
Moreover, in \S~\ref{sec:evaluation:throughput} we show that \Scheduler does not affect OpenWhisk's scalability.

%!TEX root =main.tex
%-------------------------------------------------------------------------------
\section{Implementation}
\label{sec:implementation}
%-------------------------------------------------------------------------------%

We built \Scheduler for one of the widely adopted open-source serverless platforms, Apache OpenWhisk~\cite{openwhisk}.
OpenWhisk's architecture is similar to that of a generic serverless platform as described in Section~\ref{sec:background:lifetime} allowing Hermes to be easily ported to a different platform or environment.
Moreover, the fact that OpenWhisk is used to power IBM's commercial serverless offering~\cite{ibmcloudfunctions} increases the potential impact of the performance and efficiency improvements we achieve.

%While implementing \Scheduler for a real system, we come across issues that were hidden in simulations.
%For example, there is non-negligible communication latency between the Controller and the Workers.
%Servers have a finite memory capacity while functions sharing the same server can interfere with each other.
%Finally, a real serverless platform hosts multiple different functions making cold-starts a key factor that we need to consider. 
%\Scheduler performs well both in zero-overhead simulations and higher-latency OpenWhisk deployments; this indicates that even if serverless platforms' overheads are reduced, \Scheduler will continue to improve performance. 

In this section, we first describe the details of OpenWhisk's architecture (\S~\ref{sec:implementation:openwhisk}).
Then, we describe the three different schedulers we use as baseline in our evaluation (\S~\ref{sec:implementation:scheduling}).
In \S~\ref{sec:implementation:implementation}, we discuss implementation details.

\subsection{OpenWhisk Architecture}
\label{sec:implementation:openwhisk}

%\begin{figure}[h!]
%\centering
%\includegraphics[width=0.45\textwidth]{figures/openwhisk_architecture.pdf}
%\caption[width=0.49\textwidth]{Apache OpenWhisk Architecture
%\label{fig:implementation:openwhisk}}
%\end{figure}

%\begin{figure*}[h!]
%\centering
%\includegraphics[width=0.95\textwidth]{figures/scheduling_configurations.pdf}
%\caption[width=0.99\textwidth]{The different scheduling policies implemented as baselines for Apache OpenWhisk; (a) "sticky" random load balancing used by the default OpenWhisk scheduler, (b) Late Binding, and (c) Least-Loaded.}
%\label{fig:implementation:balancing}
%\end{figure*}

Function invocations pass through a reverse proxy that terminates SSL connections and hosts a public-facing HTTP endpoint before reaching the Controller.
%The Controller is OpenWhisk's "brain".
The Controller uses a separate persistent data store, e.g., CouchDB~\cite{couchdb}, to store the system state, including the list of registered functions and their code.
Within the Controller lies a Load Balancer module which manages the scheduling of invocations to Workers.
The OpenWhisk Workers are called Invokers. 
There is usually one per server and they use containers as executors.
The Controller uses a reliable message bus, Apache Kafka~\cite{kafka}, to send function invocations to Invokers for execution.

\paragraph{Cold starts:} If a function invocation reaches an Invoker that does not have an available warm container for the specific function, we say that a cold start has occurred.
In OpenWhisk two types of cold starts can happen. 
First, it is possible that the Invoker cached the function code or executable during a previous execution of the same function.
In that case, the Invoker just needs to start a container with the function's runtime and inject the code or binary.
%It is also possible that such a container already exists; Invokers have a configurable number of pre-warmed containers for the main programming languages OpenWhisk supports (NodeJS, Go, Java, Scala, PHP, Python, Ruby, Swift, .NET, Ballerina, and Rust).
Second, function code might not be cached at the Invoker.
In that case, the Invoker fetches the function code from CouchDB and injects it to the corresponding container.

\subsection{Baseline Schedulers}
\label{sec:implementation:scheduling}

%In addition to \SchedulerT, 
We use the following schedulers as baselines:

\paragraph{OpenWhisk Scheduler ($E/LOC/PS$): }
The default OpenWhisk scheduler implements early binding, using locality-based load balancing and processor sharing at the Invokers (delegating scheduling to Linux's CFS), i.e., $E/LOC/PS$ using the notation we introduced in Section~\ref{sec:taxonomy}.
Each Invoker has a fixed capacity for invocations based on its server's available memory.
OpenWhisk' load balancer assigns invocations belonging to the same function to the same Invoker to reduce cold starts. 
A hash is calculated for every function and an Invoker is selected based on that hash.
All invocations of that function are scheduled to that Invoker.
If the Invoker selected from a function's hash is unhealthy or full, a new Invoker is randomly selected. 
This procedure is repeated until all Invokers have been checked at which point a healthy Invoker is randomly selected even if it does not have available capacity. 
If no healthy Invokers are available, the load balancer returns an error. Invocations are not queued anywhere.
%An Invoker is considered unhealthy if it produces more than 3 system errors in the previous 10 invocations it executed.
%For instance, an Invoker with 32GB of memory has a capacity of 128 invocations for 256MB functions. 

Despite the fact that this scheduler was designed explicitly for a serverless platform, it has shortcomings.
Although the locality-based balancing helps reduce cold starts in some cases (\S~\ref{sec:evaluation:coldstart}), it is unsuitable for highly-skewed workloads.
As we noted earlier, in real-world serverless workloads a small fraction of the functions accounts for most of the load (\S~\ref{sec:background:characteristics}).
This leads to server overloads and reduced performance (\S~\ref{sec:evaluation:performance}). 

\paragraph{Late Binding: } Late binding performs worse than early binding for
workloads with highly-variable execution times at high load
(\S~\ref{sec:simulation:single}).
However, since many existing scheduling systems like Sparrow and Pigeon~\cite{sparrow, pigeon} use such a policy, we implemented it for OpenWhisk to use it as a baseline in our evaluation.
Function invocations are packed to Invokers until their core capacity is reached, i.e., until the number of active invocations becomes equal to the number of cores.%(Figure~\ref{fig:implementation:balancing}).
Once this happens, additional invocations are queued at the Load Balancer.
%until an active invocation finishes execution and a core becomes available. 
This policy is core-aware~\cite{coregranular}: it requires the load balancer to take into account the number of cores in each Invoker, a departure from the original OpenWhisk scheduler model which only considers the memory capacity. 

\paragraph{Least-Loaded $(E/LL/PS)$: } Based on the simulations, the ideal scheduling policy for a workload with the characteristics of serverless functions combines early binding with least-loaded balancing and processor-sharing at the workers.
We implement such a scheduler for OpenWhisk to showcase that \Scheduler can outperform it due to improved locality without suffering from its efficiency shortcomings at low load.

\subsection{Implementation Details}
\label{sec:implementation:implementation}
The new schedulers we introduce are implemented on top of OpenWhisk public repository~\cite{openwhiskrepo}.
%Each scheduler implements OpenWhisk's \textit{LoadBalancerProvider} interface in Scala.
%The interface's basic functionality is implemented in the \textit{publish}  and \textit{processCompletion} methods.
%\textit{publish} decides on which Invoker a new function invocation is going to run.
%The method is called every time a new function invocation reaches the Controller.
%\textit{processCompletion} is called when a completion acknowledgement is received by the Controller and is used to update the load balancer's state.
For the Late Binding and Least-Loaded policies, an array is used to store Invoker load information.
For \SchedulerT, we use two such arrays, one for each mode of operation, and one map that keeps track of warm containers.
The memory overhead of these arrays is minimal; in a setup with $N$ Invokers $4 \times N$ bytes are needed for the Late Binding and Least-Loaded policies and $8 \times N$ bytes for \SchedulerT.
In our setup (described in \S~\ref{sec:evaluation:methodology}) this amounts to 32 and 64 bytes respectively.
The overhead of the map is higher as, in the worst case, it needs to have an entry for each invocation that can fit in the cluster's memory.
However, even in a large-scale deployment, this only amounts to a few megabytes.
Late Binding has the additional memory overhead of storing the queued invocations at the Controller.
Since this overhead was not a bottleneck factor in our setup and the proposed \Scheduler scheduler does not do queuing at the Controller, we do not quantify the exact memory overhead.
In OpenWhisk, all invocations and completions go through the Controller.
Hence, it is trivial to keep track of the load and the warm invocations on each Invoker.
Similarly, switching \SchedulerT's operating mode between “packing” and
“least-loaded” is simply done with an if statement without any overhead.
The three new policies are implemented in 2074 SLOC in total. 

%-------------------------------------------------------------------------------
\section{Evaluation}
\label{sec:evaluation}
%-------------------------------------------------------------------------------

We aim to answer the following questions:
\begin{enumerate}
\item How does the scheduling policy affect a workload's performance? How effective is \SchedulerT? (\S\ref{sec:evaluation:performance})
\item How does scheduling affect cold-starts? (\S\ref{sec:evaluation:coldstart})
\item What is the impact of different scheduling policies on resource consumption? (\S\ref{sec:evaluation:efficiency})
\item Is \Scheduler robust across different execution time distributions?  (\S\ref{sec:evaluation:exponential})
\item What is \SchedulerT's overhead? (\S\ref{sec:evaluation:throughput})
\end{enumerate}

\subsection{Experimental Methodology}
\label{sec:evaluation:methodology}

\begin{figure*}[t!]
\centering
\captionbox*{(a) MS Trace}
    {\includegraphics[width=.24\textwidth]{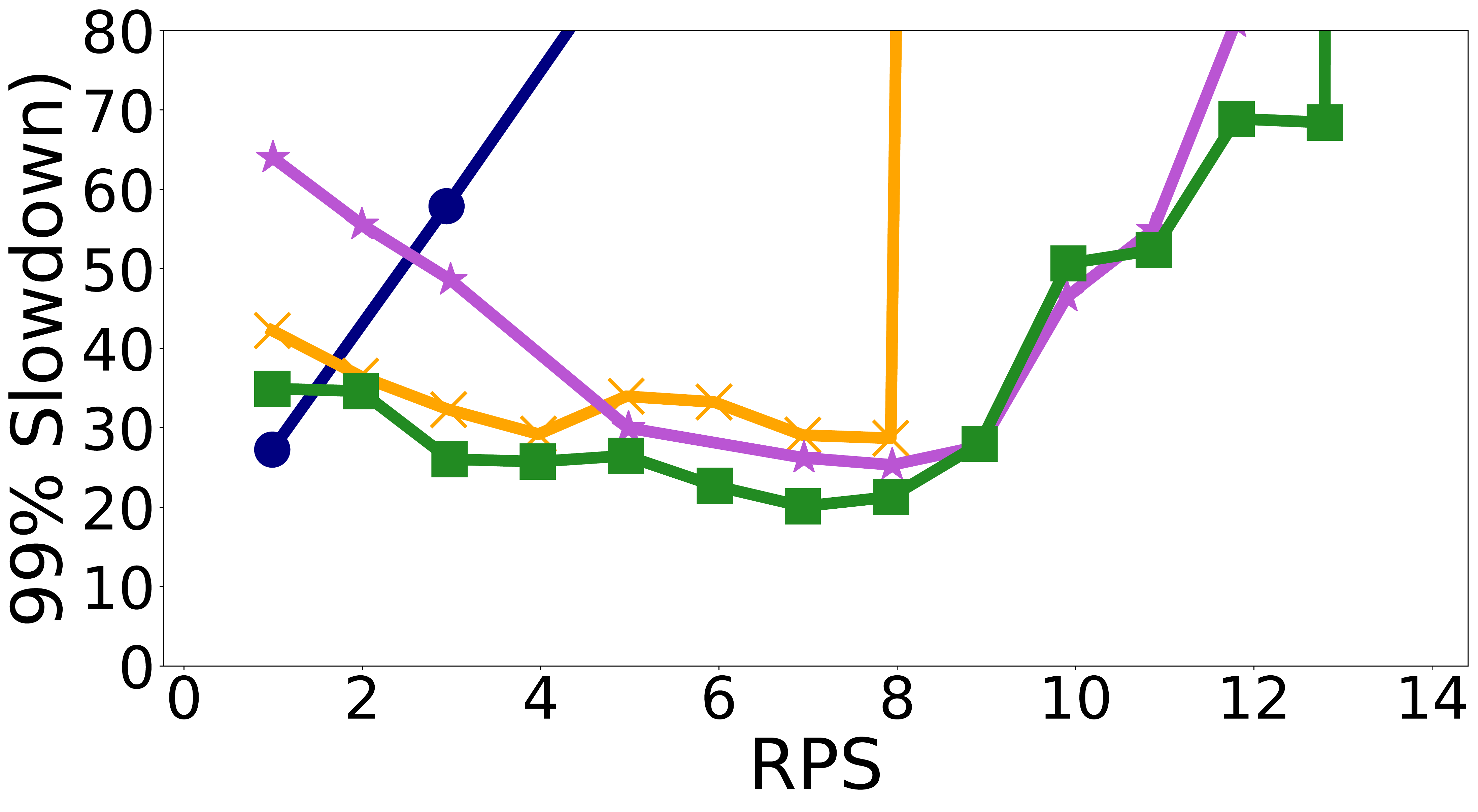}}
\captionbox*{(b) MS Representative}
{\includegraphics[width=.24\textwidth]{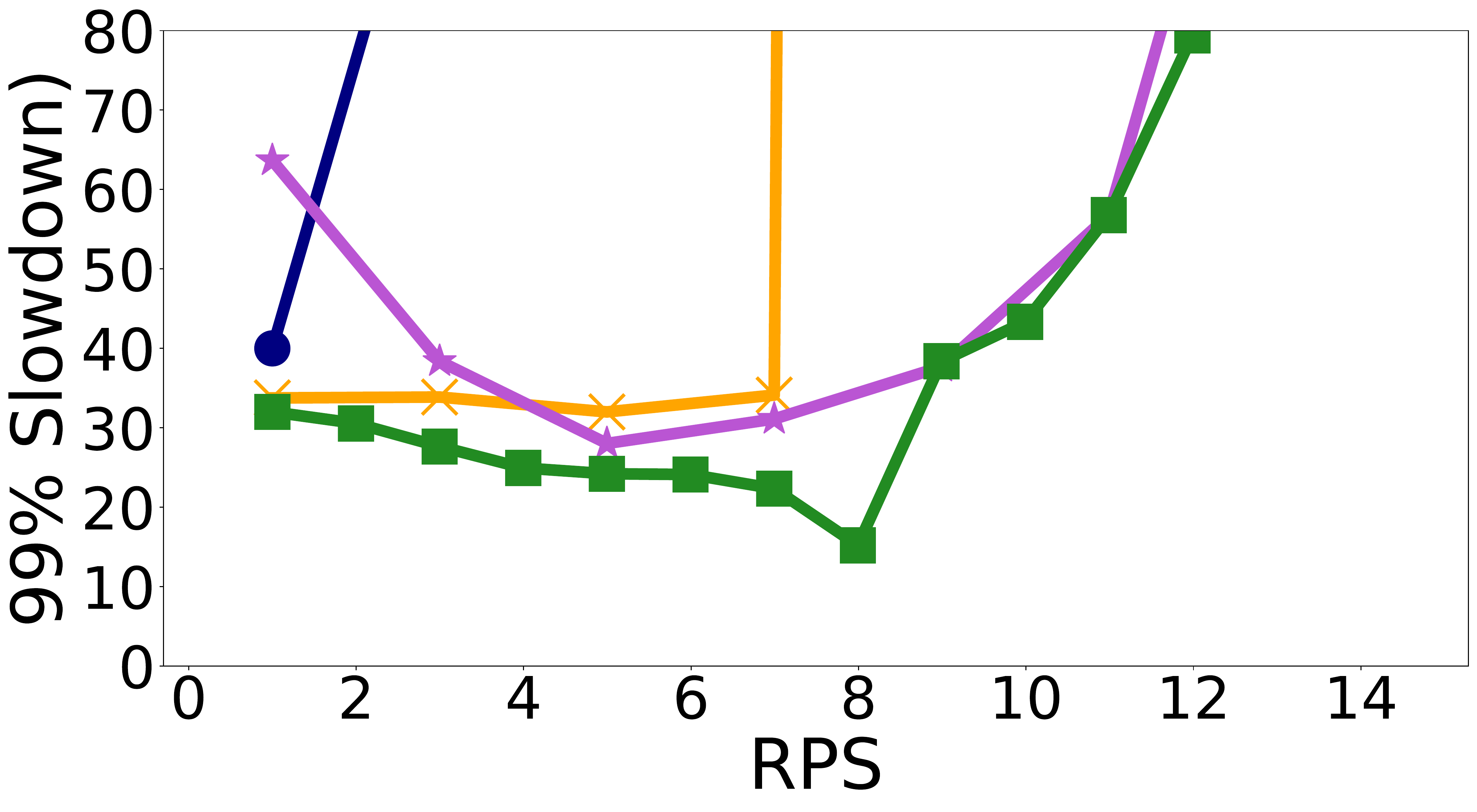}}
\captionbox*{(c) \small{Multiple-Functions-Balanced}}
{\includegraphics[width=.24\textwidth]{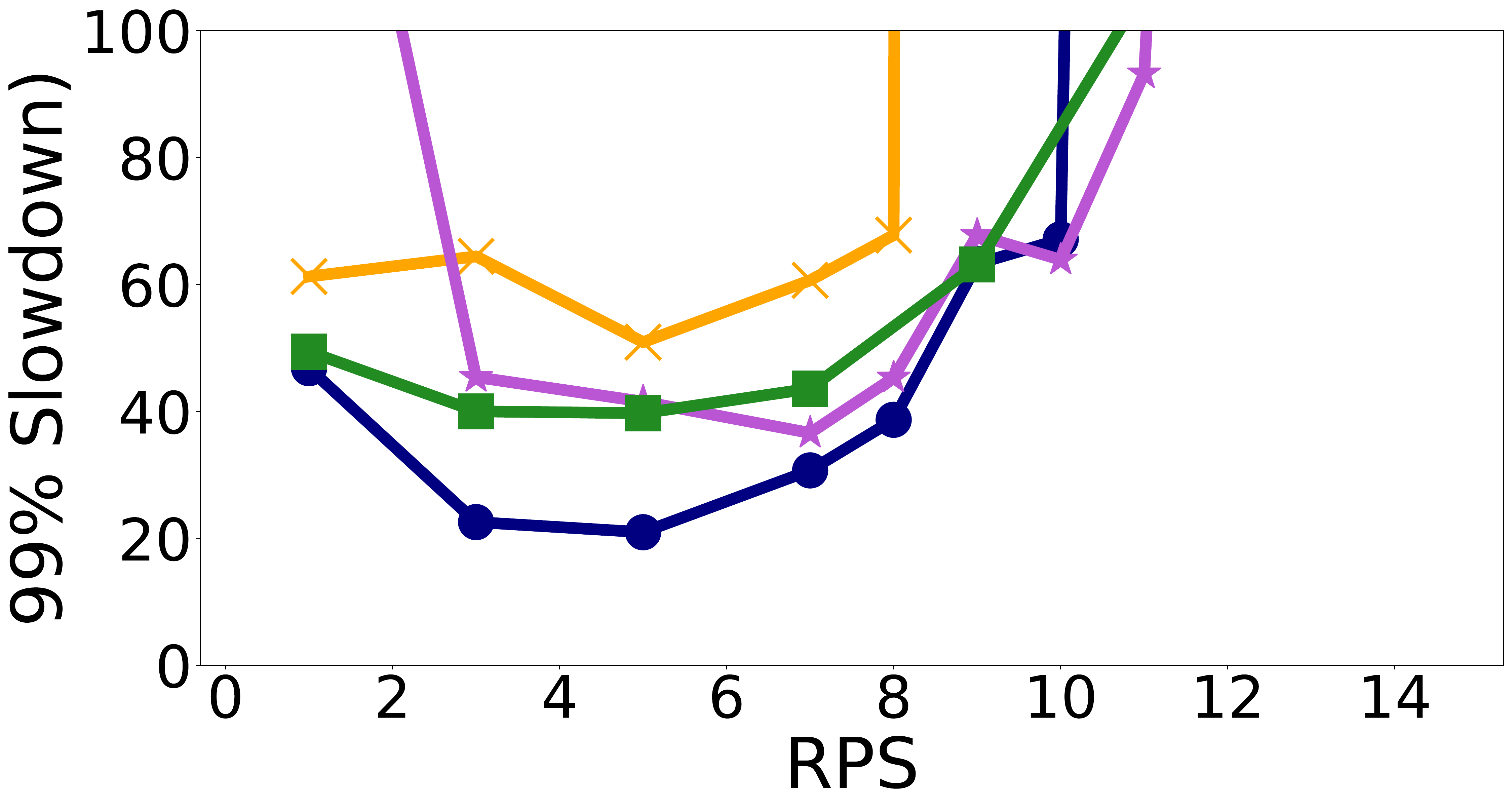}}
\captionbox*{(d) Single-Function}
    {\includegraphics[width=.24\textwidth]{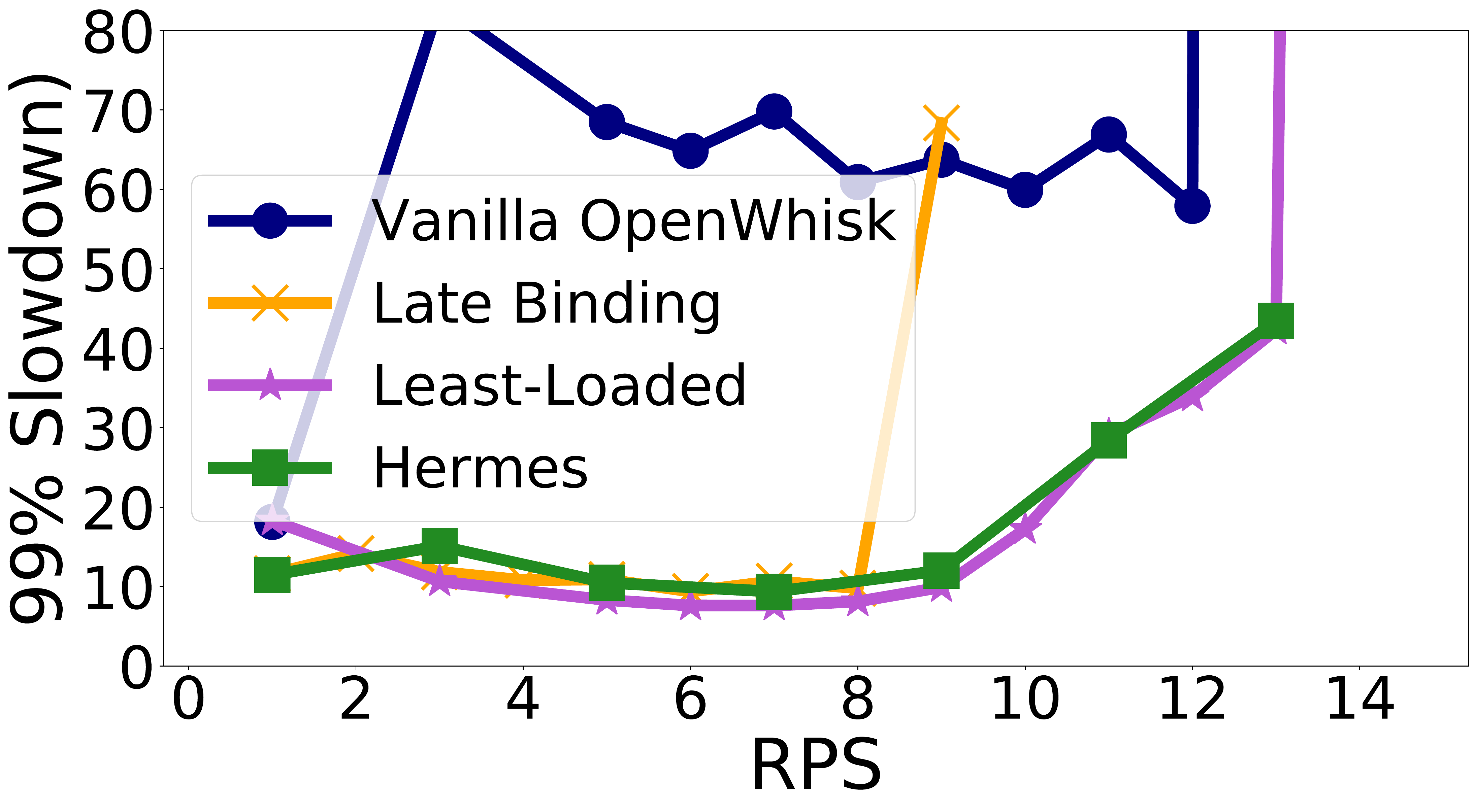}}\\
\caption{99\% slowdown as function of the load in requests per second (RPS) for four different workloads.}
\label{fig:evaluation:99_slowdown}
\end{figure*}

%\begin{figure*}[t!]
%\centering
%\captionbox*{(a) MS Trace}
%    {\includegraphics[width=.24\textwidth]{figures/lat_log_60min_50functions_unbalanced_8servers_12cores.pdf}}
%    \captionbox*{(b) \small{Multiple-Functions-Balanced}}
%    {\includegraphics[width=.24\textwidth]{figures/lat_log_60min_50functions_8servers_12cores.pdf}}
%\captionbox*{(c) Single-Function}
%    {\includegraphics[width=.24\textwidth]{figures/lat_log_60min_1function_8servers_12cores.pdf}}
%\captionbox*{(d) Twitter Trace}
%    {\includegraphics[width=.24\textwidth]{figures/lat_log_twitter_30min_50functions_unbalanced_8servers_12cores.pdf}}\\
%    \caption{99\% latency as function of the load in requests per second (RPS) for four different workloads.}
%\label{fig:evaluation:99_latency}
%\end{figure*}

%\begin{figure*}[h!]
%\centering
%\captionbox*{(a) MS Trace}
%{\includegraphics[width=.24\textwidth]{figures/median_slowdown_log_60min_50functions_unbalanced_8servers_12cores.pdf}}
%    \captionbox*{(b) \small{Multiple-Functions-Balanced}}
%{\includegraphics[width=.24\textwidth]{figures/median_slowdown_log_60min_50functions_8servers_12cores.pdf}}
%\captionbox*{(c) Single-Function}
%{\includegraphics[width=.24\textwidth]{figures/median_slowdown_log_60min_1function_8servers_12cores.pdf}}
%\captionbox*{(d) Twitter Trace}
%    {\includegraphics[width=.24\textwidth]{figures/median_slowdown_log_twitter_30min_50functions_unbalanced_8servers_12cores.pdf}}\\
%    \caption{50\% slowdown as function of the load in requests per second (RPS) for
%    four different workloads.}
%\label{fig:evaluation:50_slowdown}
%\end{figure*}

\begin{figure*}[ht!]
\centering
\captionbox*{(a) MS Trace}
    {\includegraphics[width=.24\textwidth]{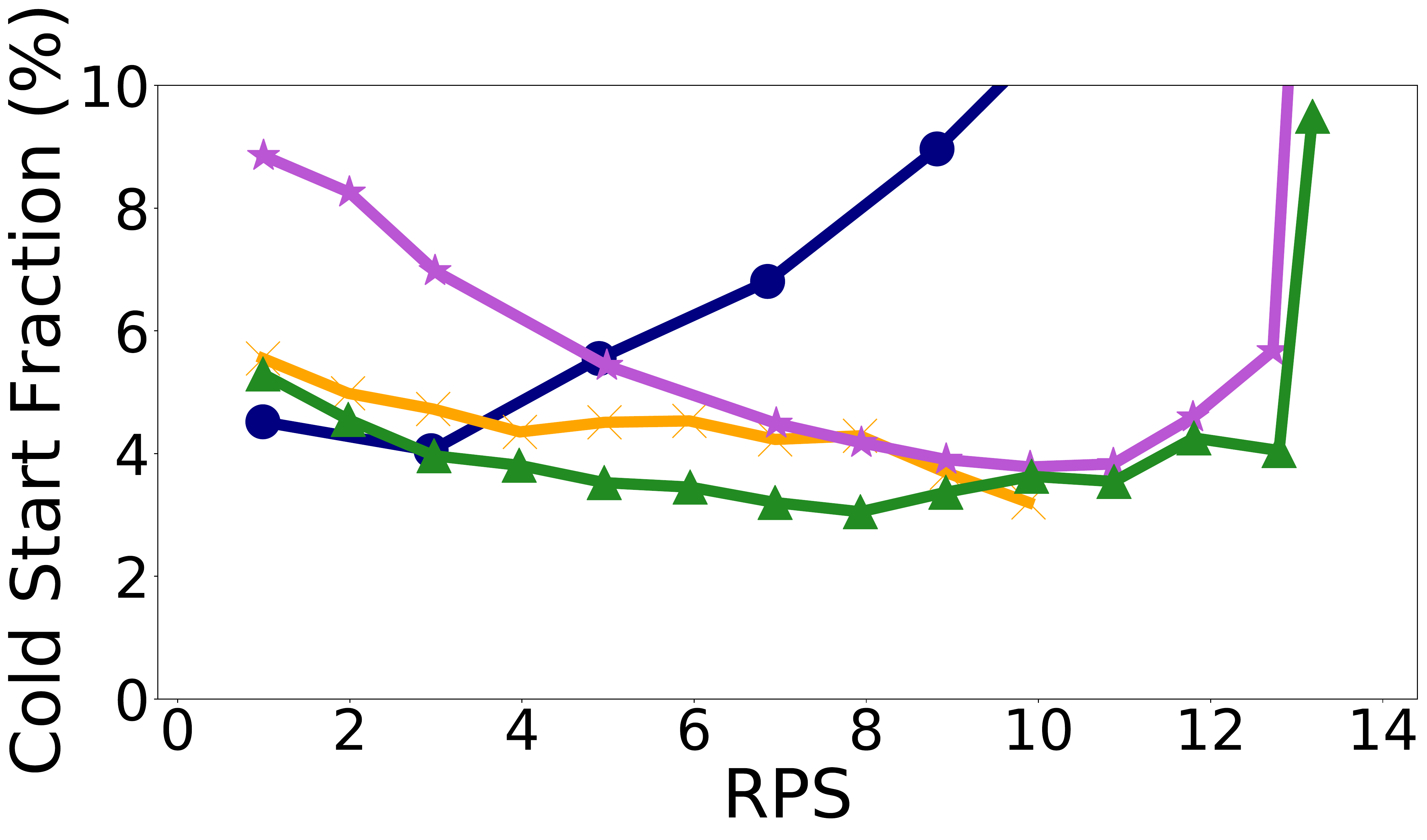}}
\captionbox*{(b) MS Representative}
{\includegraphics[width=.24\textwidth]{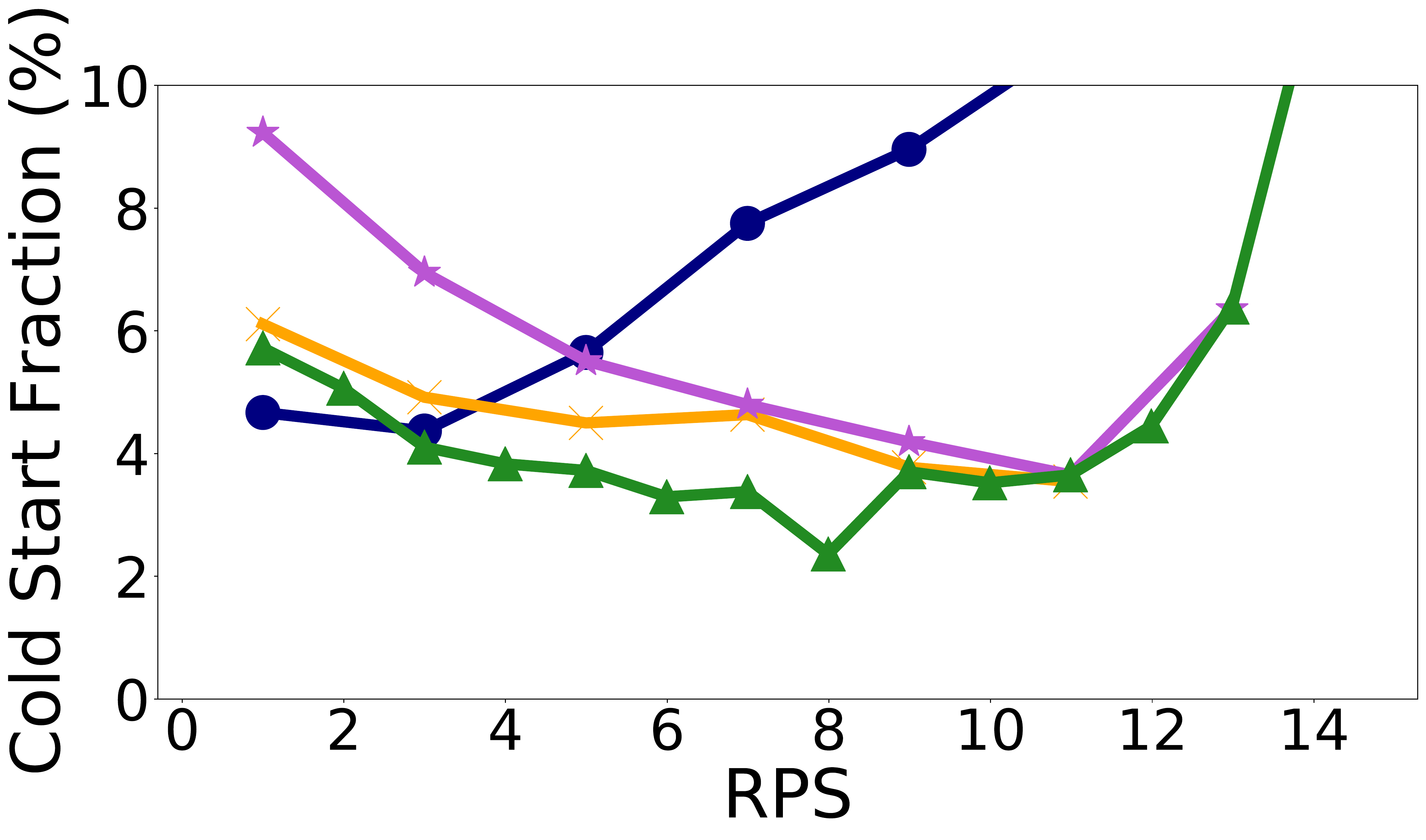}}
\captionbox*{(c) \small{Multiple-Functions-Balanced}}
{\includegraphics[width=.24\textwidth]{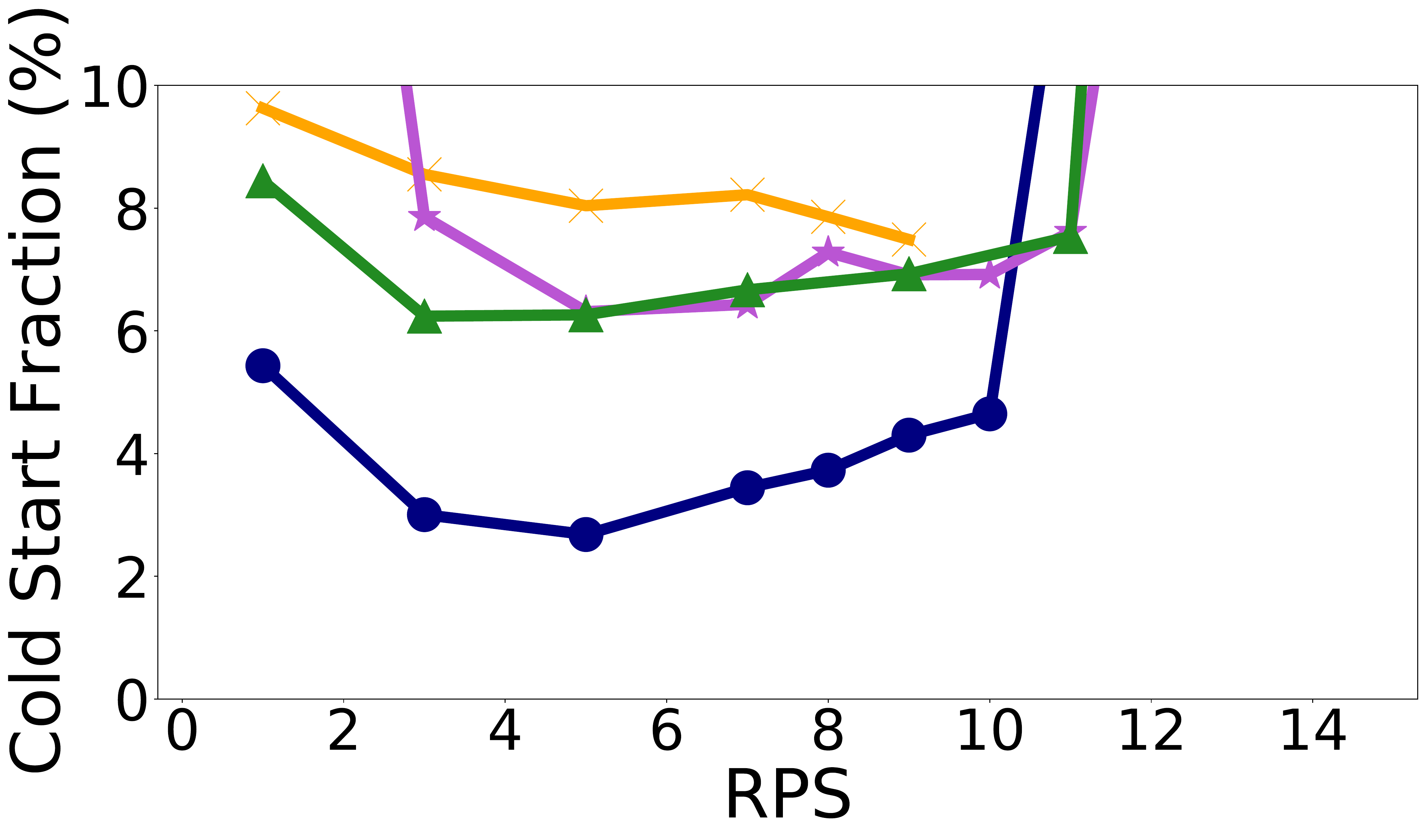}}
\captionbox*{(d) Single-Function}
    {\includegraphics[width=.24\textwidth]{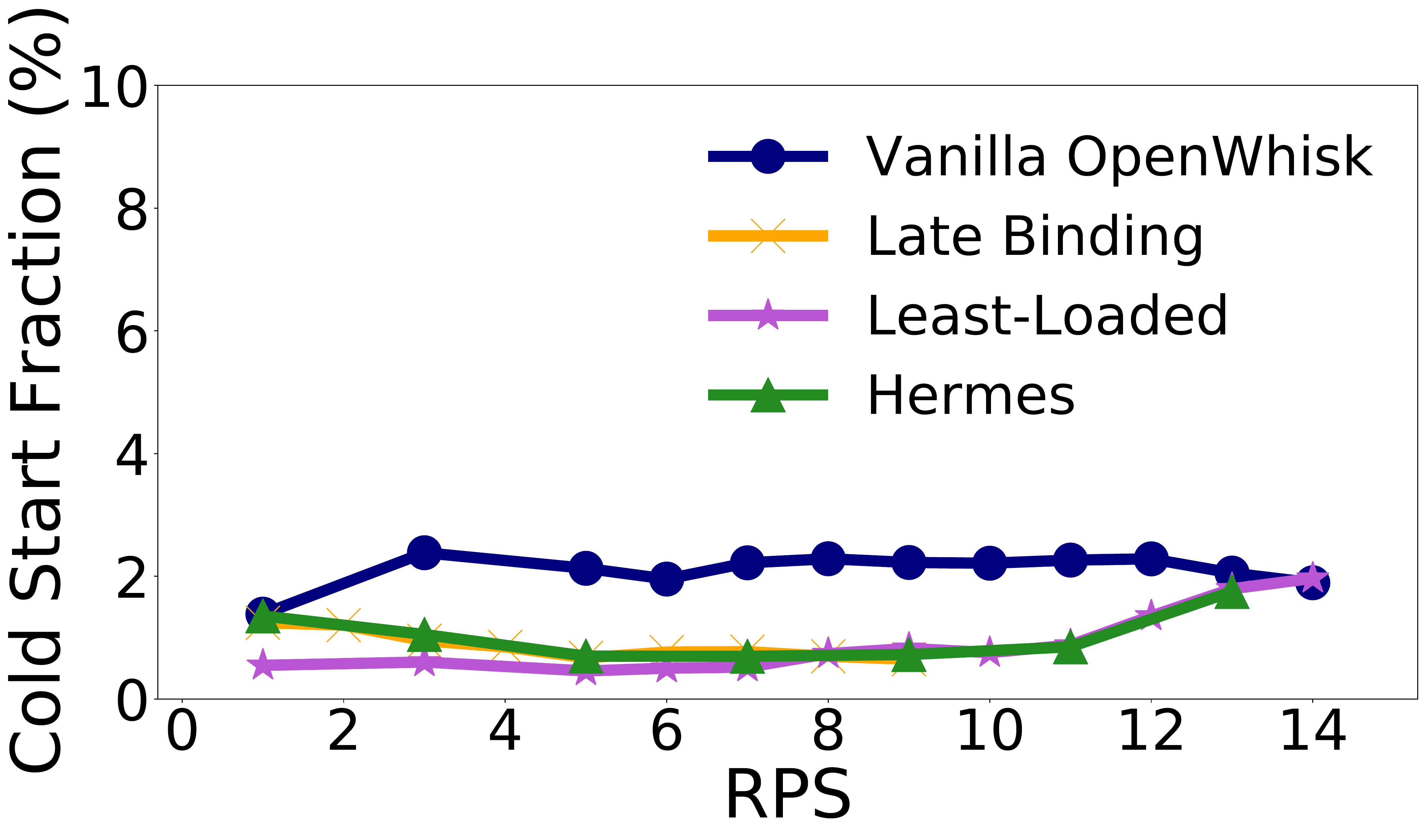}}\\
\caption{\% of cold start function invocations as a function of the load in requests per second (RPS).}
\label{fig:evaluation:cold_start}
\end{figure*}

\paragraph{Experimental setup: } We use a cluster of 9 machines, connected through an Arista 7050-S switch with 48 10GbE ports.
All machines have one Intel Xeon E5-2630 CPU operating at 2.3GHz running Ubuntu LTS 16.0.4 with the 4.4.0 Linux kernel and 32GB memory capacity.
Each CPU has 12 cores.
Hyperthreading is disabled to improve performance predictability.
One machine hosts the OpenWhisk Controller and CouchDB while the rest of the machines host an Invoker each.
Each Invoker can use up to 26,624MB, defined through the \textit{userMemory} OpenWhisk parameter.
%Our setup has 96 Invoker cores in total, which is more than 2.5$\times$ the compute capacity used for the evaluation in~\cite{msrserverless}.
We opted to do the evaluation in a private cluster to avoid the performance variability associated with the public cloud and focus only on the effect of scheduling on each workload's performance.
%We are confident our findings will translate to larger deployments since the
%proposed \Scheduler scheduler introduces no additional overhead compared with
%the vanilla OpenWhisk scheduler (Section~\ref{sec:evaluation:overhead}).
Similar to existing related work~\cite{msrserverless, archimpl}, we use FaasProfiler as the load generator.
%We modify FaasProfiler to collect runtime metrics across different servers that host OpenWhisk Invokers.
The experiment for each data point runs for 1 hour.
%To avoid skewing of the results during the start-up or tear-down phases, we collect slowdown metrics only for the invocations that started in the 25-35 minutes window.

\noindent\textbf{Baselines: } We compare \Scheduler with the following schedulers: (i) Vanilla OpenWhisk Scheduler, (ii) Late Binding Scheduler, (iii) Least-Loaded Scheduler.

\noindent\textbf{Workloads:} 
To evaluate the performance of \SchedulerT, we use five workloads, two derived
from production traces, and three that test emerging serverless use cases and extreme scenarios. 

First, we use \textbf{MS Trace}, a workload directly derived from the Azure production trace~\cite{msrserverless}.
The trace includes information about function invocations across Microsoft's data centers and tens of thousands of different functions.
To generate this workload, we use the same methodology that the authors of the paper that accompanied the trace release used for their analysis:
(a) We select 50 different functions from the trace and scale the number of invocations to produce different load levels.
To model the highly skewed function invocations found in the original trace, we randomly select one function of high popularity and 49 functions of medium popularity.
Since the trace does not specify the implementation language of each function, we opted to use Javascript due to its prevalence in serverless computing~\cite{stateofserverless}.
Using a different language would not alter our conclusions since functions written in different languages face similar overheads~\cite{peeking}.
(b) We use a Log-normal distribution with $\mu=-0.38$ and $\sigma=2.36$ for function execution times similar to the original trace.
Invocations across all functions follow this distribution, capturing
variability in running times even within a function.
Similar characteristics also appear in Amazon's serverless
offering~\cite{stateofserverless}, enhancing the significance and the validity of
this workload.

We also need to ensure that workloads with different characteristics that
might emerge in the future perform well under \SchedulerT.
To showcase \SchedulerT's robustness, we consider additional workloads
based on different scenarios across three axes: arrival pattern, skew, and
execution time distribution.
\begin{itemize}[topsep=0pt]
    \item \textbf{Arrival Pattern:} To avoid making arbitrary low-level assumptions regarding missing data in the Microsoft production trace, such as how function arrivals are distributed within each minute, we use open-loop Poisson processes to model function invocation arrivals, following the standard practice in existing relevant research~\cite{treadmill}.
        To model the original trace's skew, we use again 50 different functions, one accounting for 90\% of the overall load while the rest equally account for the remaining 10\%.
        We  use the execution time distribution from the MS Trace.
        We refer to this workload as \textbf{MS Representative}.
    \item \textbf{Skew:} The \textit{MS Trace} models the case of a
        highly skewed workload. We also evaluate the performance of \Scheduler
        in two very different scenarios. First, in the \textbf{Single-Function}
        workload, all function invocations belong to a single function. This
        pattern is characteristic of analytics applications where
        many functions are spun up to do the same type of
        processing~\cite{excamera, gg, sprocket, pocket, pywren, infinicache}
        and showcases \SchedulerT's performance under extreme skew.
        Second, for the \textbf{Multiple-Functions-Balanced} workload, we
        deploy 50 different functions, each equally contributing to the overall
        load. This pattern is characteristic of an entirely homogeneous
        workload with zero skew. We use 
        the execution time and interarrival time distributions from the MS
        Representative workload for both of these workloads.
    \item \textbf{Execution Time Distribution:} As mentioned before,
        \textit{MS Trace}'s invocations have extremely high execution time
        variability. We want to ensure that \Scheduler provides good
        performance for workloads with more homogeneous execution times.
        To validate that, we create a \textbf{Homogeneous\-/Execution\-/Times}
        workload with the skew and inter-arrival distribution of the MS Trace
        workload but with execution time that
        follows a light-tailed exponential distribution. We set the mean to 8.9
        seconds, similar to the Log-normal distribution of the MS
        Trace workload. The performance of this workload is evaluated in
        \S~\ref{sec:evaluation:exponential}.
\end{itemize}

We set the memory used by each function to 256MB across all workloads.
Hence, each Invoker has a capacity of up to $26624 / 256 = 104$ function invocations.

\subsection{Performance Analysis}
\label{sec:evaluation:performance}

Figure~\ref{fig:evaluation:99_slowdown} compares the 99\% slowdown of four different workloads under the four different schedulers; we make the following observations.
First, similar to what we observed in the simulations, Late Binding and the Vanilla OpenWhisk schedulers can support a lower load than the rest of the schedulers.
For Late Binding, the 99\% slowdown explodes for a 39\% lower load (8 vs. 13 RPS) than
the Least-Loaded and \Scheduler schedulers for the MS Trace, MS Representative, and Single-Function workloads.
For Vanilla OpenWhisk, the 99\% slowdown explodes from very low loads due to the skew.
Second, \Scheduler performs better - providing up to 50\% lower slowdown - at low and medium load compared to the locality-unaware Least-Loaded scheduler.
Third, for the Multiple-Functions-Balanced workload, we observe a different behavior. 
Each function equally contributes to the overall load allowing the locality-based balancing used by Vanilla OpenWhisk to distribute the load equally among the Invokers and achieve 99\% slowdown better than the other policies at low load.

The 50\% slowdown shows similar characteristics to those of the 99\% slowdown.
We also observed that, as the simulations showed (Figure~\ref{fig:scheduling:basic}a), the different schedulers perform similarly in terms of tail latency.
We omit figures for these two metrics due to space constraints.

\noindent\textbf{Conclusion: } The observations confirm the conclusions drawn by our simulations that the least-loaded policy used by \Scheduler can support higher load both than Late Binding and pure locality-based approaches (Vanilla OpenWhisk).
Moreover, the fact that \Scheduler is locality-aware allows it to achieve a lower slowdown than pure load-aware scheduling.
\Scheduler retains these performance gains even for improbable homogeneous workloads, sacrificing little slowdown.
We also show that the Vanilla OpenWhisk scheduler is optimized for the "wrong" workload, i.e., a completely homogeneous mix of functions that does not appear in practice.

\subsection{Cold Start Analysis}
\label{sec:evaluation:coldstart}

Figure~\ref{fig:evaluation:cold_start} presents the percentage of function invocations that led to a cold start for the four workloads.
\Scheduler triggers very few cold starts for the MS Trace and MS Representative workloads as it steers invocations to Invokers with warm containers when possible.
The Least-Loaded scheduler has a particularly high cold start rate at low and medium load: containers for all 50 different functions need to be started in all 8 Invokers.
The higher the load, the fewer the cold starts under the Late Binding scheduler since it is more likely that an invocation is first queued at the Controller and hence is scheduled to a warm container finishing the execution of a previous invocation.

The Vanilla OpenWhisk scheduler generates more cold starts for the Single-Function, MS Trace, and MS Representative workloads because it tends to overload individual Invokers.
An overloaded Invoker is less likely to have a warm container available; a new one often needs to be created.
However, validating our slowdown measurements, Vanilla OpenWhisk has significantly fewer cold starts than the other schedulers for the Multiple-Functions-Balanced workload due to its sticky load balancing.
Invocations of the same function are more likely to be scheduled to the same Invoker finding a warm container.

\noindent\textbf{Conclusion: } \SchedulerT, being both locality- and load-aware, causes fewer cold starts, explaining the lower slowdown shown in Figure~\ref{fig:evaluation:99_slowdown}.
The use of predictive policies~\cite{msrserverless,ofc}, which are orthogonal to \SchedulerT, could reduce cold starts even more.

\subsection{Resource Consumption}
\label{sec:evaluation:efficiency}

\begin{figure}[h!]
\centering
{\includegraphics[width=.23\textwidth]{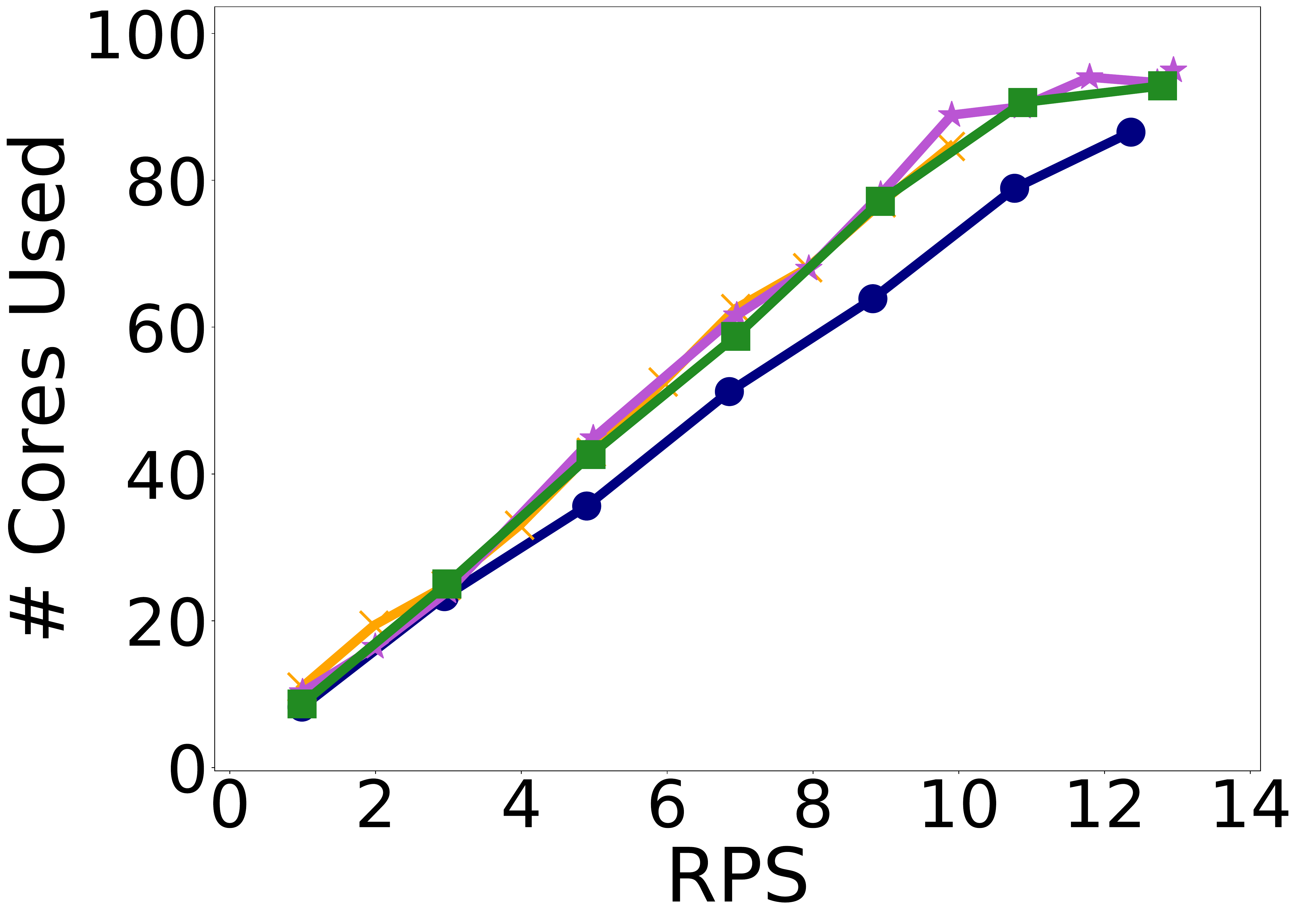}}
{\includegraphics[width=.23\textwidth]{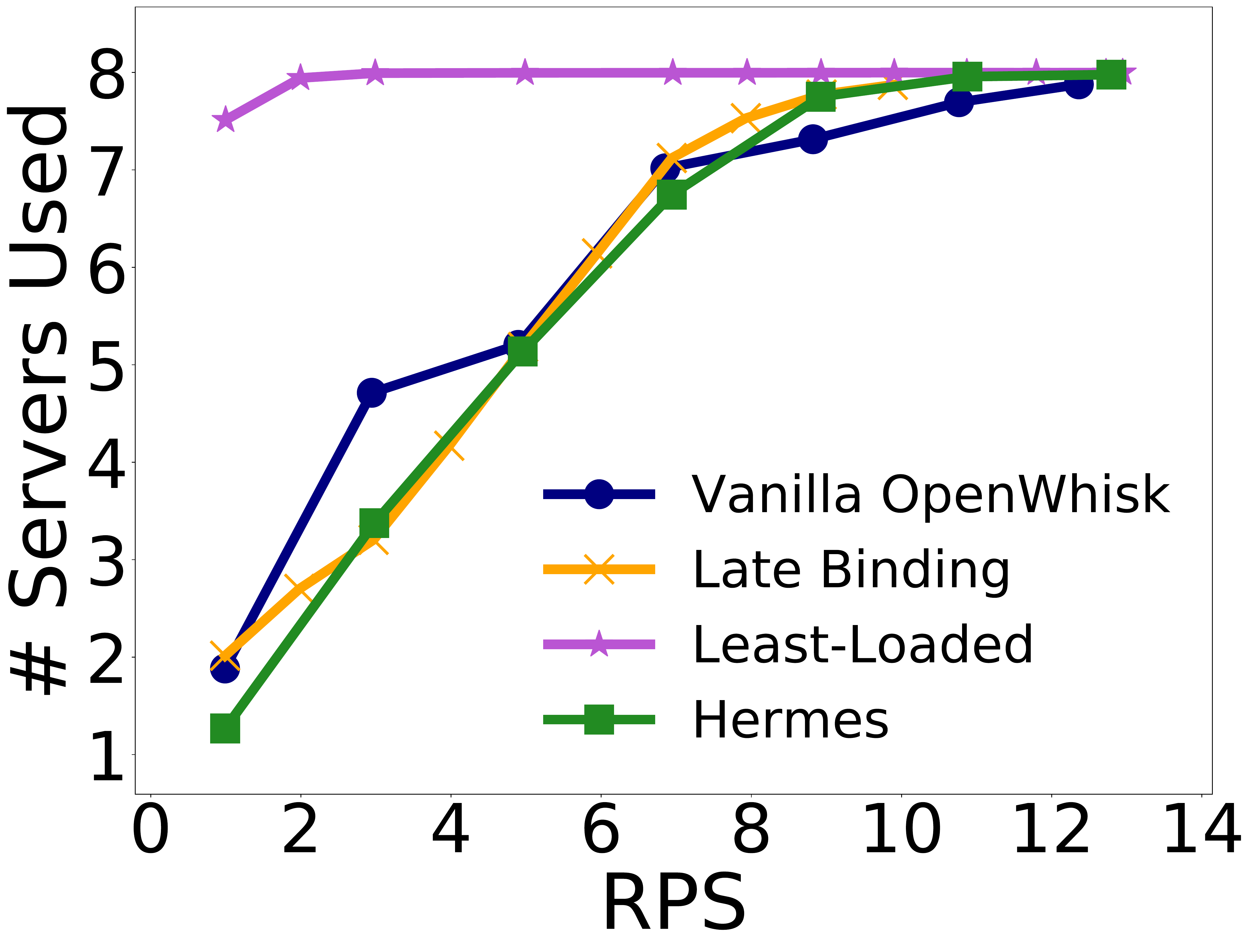}}\\
\caption{Average \# Cores and \# Servers utilized during the execution of the
    MS Trace workload.}
\label{fig:evaluation:util}
\end{figure}

We present the average resource utilization when using different schedulers for the MS Trace workload (Figure~\ref{fig:evaluation:util}).
For this experiment, we collect utilization metrics every second and consider a
server to be utilized if a serverless function uses it over the 1-second
monitoring period.
We omit the results for the other workloads since they are almost identical.

The number of cores used is similar across policies.
However, when we consider server utilization, the advantages of \Scheduler are apparent.
It uses fewer servers than the Least-Loaded scheduler at low load by packing invocations, while it achieves lower slowdown.
The other two schedulers (Late Binding and Vanilla OpenWhisk) use about the same number of servers as \Scheduler while providing worse performance.

\noindent\textbf{Conclusion: } \Scheduler achieves a lower slowdown than the Least-Loaded scheduler while providing better consolidation.

\subsection{Robustness to Different Execution Time Distributions}
\label{sec:evaluation:exponential}

\begin{figure}[h!]
\centering
\captionbox*{(a) 50\% Slowdown}
    {\includegraphics[width=.23\textwidth]{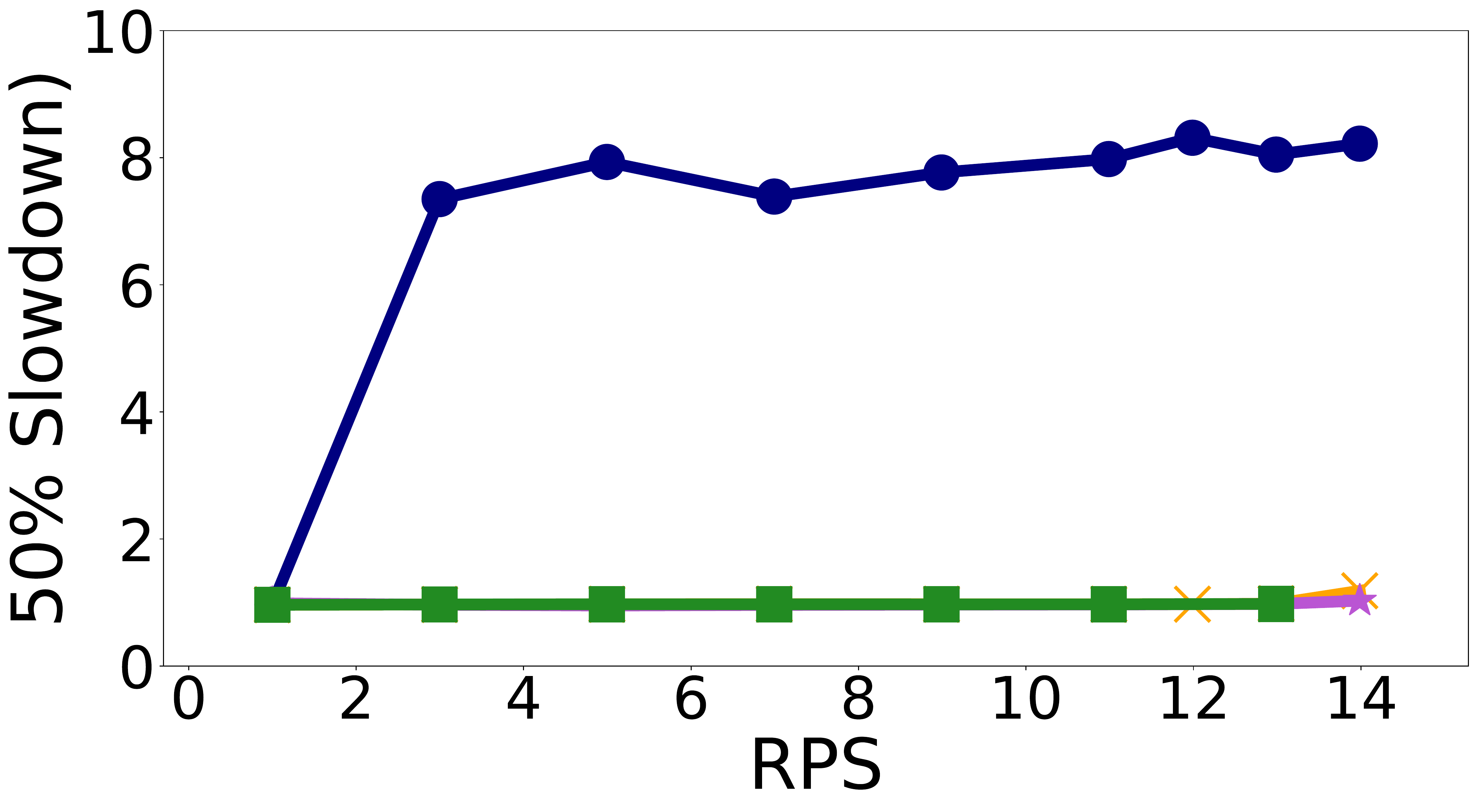}}
\captionbox*{(b) 99\% Slowdown}
    {\includegraphics[width=.23\textwidth]{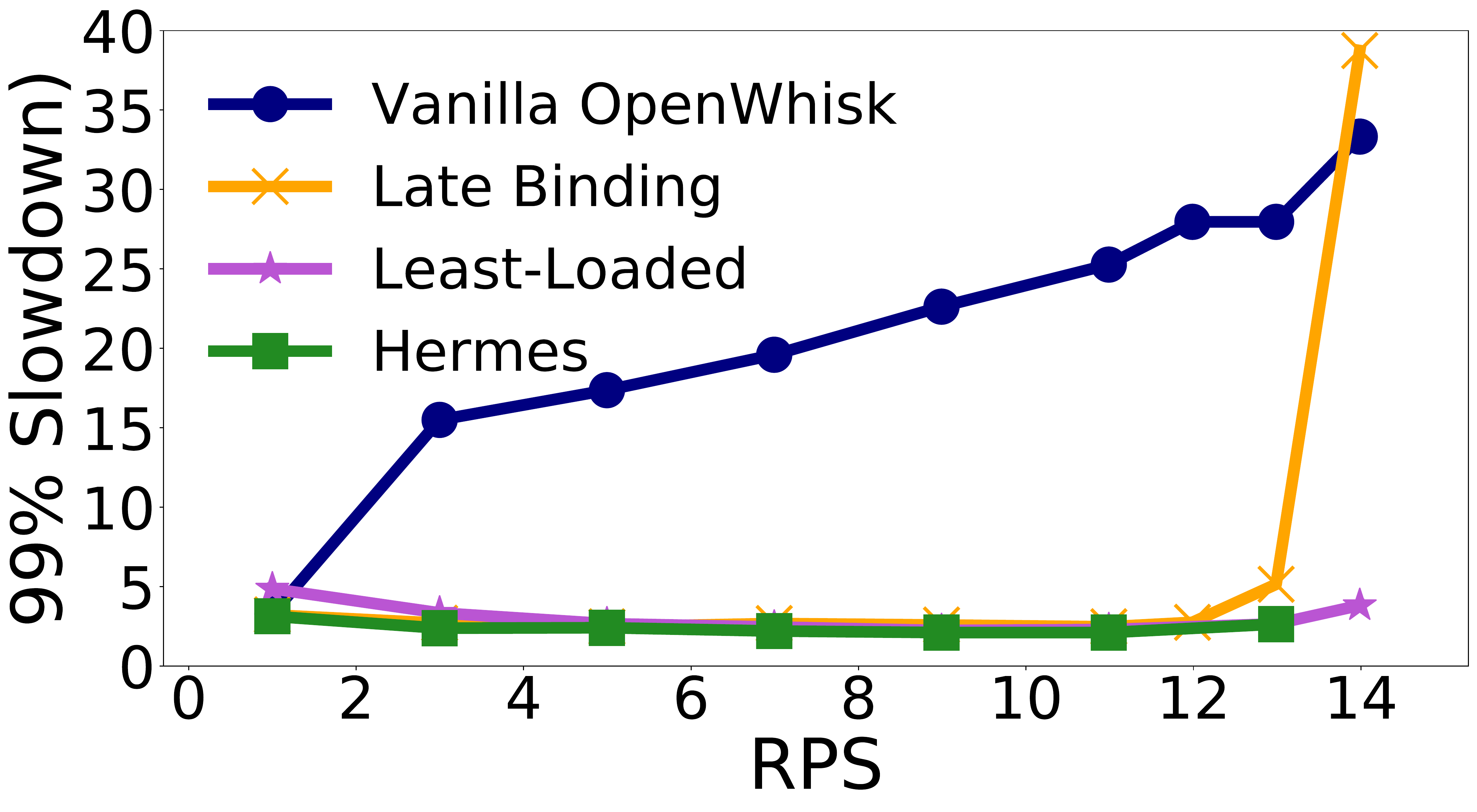}}\\
\caption{Slowdown as a function of the load for the Homogeneous-Execution-Times workload.}
\label{fig:evaluation:exponential}
\end{figure}

In Figure~\ref{fig:evaluation:exponential}, we observe that the slowdown for
the \textit{Homogeneous\-/Execution\-/Times} workload under \Scheduler - both at the median and the tail - is very similar to the slowdown under the Least-Loaded and Late Binding policies.
The Vanilla OpenWhisk scheduler again causes a higher slowdown at low load due to
skewed function load and sticky load balancing.

\textbf{Conclusion: } \Scheduler matches or exceeds the performance achieved by
existing schedulers even for workloads with homogeneous execution times,
demonstrating its robustness.

\subsection{OpenWhisk's Overheads}
\label{sec:evaluation:throughput}
In this section, we examine whether OpenWhisk's inherent overheads affect our conclusions regarding \SchedulerT.
The table below shows each scheduler's highest throughput, i.e., the maximum number of zero-work
function invocations completed in one second. In this scenario, the
Controller/Scheduler is the bottleneck since the Invokers never exceed 40\%
utilization. The throughput is similar for all schedulers showing
that \Scheduler does not cause any additional overhead.
\begin{center}
\begin{tabular}{ |c||c| }
\hline
    Scheduler & Throughput (RPS) $\pm$ std \\
\hline
\hline
    Vanilla OpenWhisk & $3833.8 \pm 43.8$ \\
    Late Binding & $3830.2 \pm 93.4$ \\
    Least-Loaded & $3832.4 \pm 58.0$ \\
    \SchedulerT & $3818.4 \pm 32.0$ \\
\hline
\end{tabular}
\end{center}

Moreover, we see that the throughput achieved in the rest of our more realistic experiments is much smaller than the maximum one achieved by all schedulers.
Hence, any difference in performance we observe between the different schedulers results from the scheduling policies they use rather than the mechanisms they employ.

We also examine the effect of OpenWhisk's overheads on the slowdown.
When the system operates under low load, the median slowdown is approximately 1, i.e., the minimum possible.
However, the 99\% slowdown is 20-30 even for low load, as shown in Figure~\ref{fig:evaluation:99_slowdown}.
This is an artifact of OpenWhisk's overheads, as even a warm invocation still needs to go through an HTTP server, trigger a CouchDB look-up, go through the Load Balancer and Apache Kafka, and finally reach an Invoker.
The scheduling decision takes less than 0.5msec, contributing little to the overall overhead.
More efficient mechanisms (e.g., bespoke communication layer instead of Kafka, in-memory data structures instead of CouchDB, etc.) are orthogonal to our work as they could reduce the 99\% slowdown across all schedulers, but they would not change each worker's load and thus the shape of the curve.

%-------------------------------------------------------------------------------
\section{Conclusion}
This paper presents \SchedulerT, a scheduler that caters to the unique characteristics of serverless functions and achieves near optimal performance.
\Scheduler uses three key techniques: early binding, hybrid load balancing, i.e., consolidation at low load and least-loaded balancing at high load, and processor sharing scheduling at the individual workers.
Implementing \Scheduler for Apache OpenWhisk we demonstrate that it can provide up to 85\% lower slowdown and support up to 60\% higher load compared to existing approaches for a real-world production workload.
Additional experiments demonstrate that \Scheduler is robust to workloads with different function mixes and execution time distributions.
Hence, we believe that \Scheduler provides an answer to how scheduling for serverless functions should be done.
%-------------------------------------------------------------------------------

\bibliographystyle{plain}
\bibliography{paper}

\end{document}